\documentclass{aa}
\usepackage{graphicx}
\usepackage{txfonts}
\usepackage{amsmath,amssymb}
\usepackage{mathrsfs}	
\usepackage{mathtools}  
\usepackage{color}
\usepackage{colortbl}
\definecolor{Gray}{gray}{0.9}
\newcolumntype{a}{>{\columncolor{Gray}}l}
\newcolumntype{b}{>{\columncolor{Gray}}r}
\usepackage{array}

\usepackage[bookmarks=true]{hyperref}
\hypersetup{
  colorlinks=true,
  pdfborder=2 2 0.,
  linkcolor=red,
  anchorcolor=black,
  citecolor=blue,
  urlcolor=black,
}

\usepackage{natbib}
\bibpunct{(}{)}{;}{a}{}{,} 

\newcommand{\Msun}{M_\odot}

\newcommand{\Lsun}{L_\odot}
\newcommand{\Mstar}{M_\star}

\newcommand{\fstar}{f_\star}

\newcommand{\fb}{f_{\rm b}}

\newcommand{\de}{{\rm d}}
\newcommand{\Mh}{M_{\rm h}}

\newcommand{\Jr}{J_r}

\newcommand{\bJ}{{\bf J}}
\newcommand{\btheta}{\boldsymbol{\theta}}
\newcommand{\bx}{{\bf x}}
\newcommand{\bv}{{\bf v}}
\newcommand{\Vesc}{V_{\rm esc}}
\newcommand{\Vrad}{V_{\rm los}}
\newcommand{\Vlos}{V_{\rm los}}
\newcommand{\Vcirc}{V_{\rm circ}}
\newcommand{\sigmalos}{\sigma_{\rm los}}
\newcommand{\Vh}{V_{\rm h}}
\newcommand{\Vgal}{V_{\rm gal}}
\newcommand{\Vflat}{V_{\rm flat}}
\newcommand{\MGCS}{M_{\rm GCS}}
\newcommand{\LK}{L_{\rm K}}
\newcommand{\sigFJ}{\sigma_{\star,\,\rm 1kpc}}
\newcommand{\cprime}{c'}

\defcitealias{PFM19}{PFM19}

\begin{document}

\title{Dynamical evidence for a morphology-dependent relation between the stellar and halo masses
       of galaxies}
\titlerunning{Morphology-dependent stellar-to-halo mass relation}
\authorrunning{L. Posti \& S. M. Fall}

  \author{Lorenzo Posti\inst{1}\fnmsep\thanks{lorenzo.posti@astro.unistra.fr}
           \and
          S. Michael Fall\inst{2}
          }
  \institute{Universit\'e de Strasbourg, CNRS UMR 7550, Observatoire astronomique de Strasbourg,
             11 rue de l'Universit\'e, 67000 Strasbourg, France.
            \and
            Space Telescope Science Institute, 3700 San Martin Drive, Baltimore, MD 21218, USA.
             }
      \date{Received XXX; accepted YYY}

  \abstract{
            We derive the stellar-to-halo mass relation (SHMR), namely $\fstar \propto \Mstar/\Mh$
            versus $\Mstar$ and $\Mh$, for early-type galaxies from their near-infrared luminosities
            (for $\Mstar$) and the position-velocity distributions of their globular cluster systems
            (for $\Mh$). Our individual estimates of $\Mh$ are based on fitting a flexible dynamical
            model with a distribution function expressed in terms of action-angle variables and
            imposing a prior on $\Mh$ from the correlation between halo concentration and mass in the
            standard $\Lambda$ Cold Dark Matter ($\Lambda$CDM) cosmology. We find that the SHMR for
            early-type galaxies declines with mass beyond a peak at $\Mstar\sim 5\times 10^{10}\Msun$
            and $\Mh \sim 1 \times 10^{12} \Msun$ (near the mass of the Milky Way). This result is
            consistent with the standard SHMR derived by abundance matching for the general population
            of galaxies, and with previous, less robust derivations of the SHMR for early-type
            galaxies. However, it contrasts sharply with the monotonically rising SHMR for late-type
            galaxies derived from extended HI rotation curves and the same $\Lambda$CDM prior on
            $\Mh$ as we adopt for early-type galaxies. We show that the SHMR for massive galaxies
            varies more or less continuously, from rising to falling, with decreasing disc
            fraction and decreasing Hubble type. We also show that the different SHMRs for late-type
            and early-type galaxies are consistent with the similar scaling relations between their
            stellar velocities and masses (the Tully-Fisher and the Faber-Jackson relations). As we
            demonstrate explicitly, differences in the relations between the stellar and halo virial
            velocities account for the similarity of the scaling relations. We argue that all these
            empirical findings are natural consequences of a picture in which galactic discs are built
            mainly by relatively smooth and gradual inflow, regulated by feedback from young stars,
            while galactic spheroids are built by a cooperation between merging, black-hole fuelling,
            and feedback from active galactic nuclei.
            }
  \maketitle

\section{Introduction} \label{sec:intro}

Galaxies consist of stars and interstellar gas in relatively compact bodies surrounded by more
extended halos of dark matter and circumgalactic gas. The composition of the dark matter
is unknown, but it is believed to be elementary particles that interact only gravitationally
with baryons. In the standard $\Lambda$ Cold Dark Matter ($\Lambda$CDM) paradigm, the
assembly of galactic halos by gravitational clustering is relatively simple and well understood,
while the inflow and outflow of gas and the formation of stars by both gravitational and
hydrodynamical processes are much more complex and are topics of intense current research. One
of the most useful empirical constraints in these studies -- and the focus of this paper -- is
the ratio
\begin{equation} \label{eq:fstar}
    \fstar \equiv \frac{\Mstar}{\fb\Mh}
\end{equation}
of the mass in stars $\Mstar$ to that in dark matter $\Mh$ within a galaxy normalised by the
cosmic baryon fraction $\fb$. This ratio represents a sort of global star formation efficiency,
averaged over space and time, for that galaxy.

The variation of $\fstar$ with $\Mstar$ or $\Mh$ is called the stellar-to-halo mass relation
(SHMR). This has now been derived using several different techniques: abundance matching
\citep{ValeOstriker04,Conroy+06,Behroozi+13,Moster+13}, halo occupation distributions
\citep{PeacockSmith00,Kravtsov+04,Reddick+13}, group catalogues \citep{Zheng+07,Yang+08},
weak galaxy-galaxy lensing \citep{Leauthaud+12,vanUitert+16}, galaxy clustering
\citep{ZuMandelbaum15,Tinker+17}, empirical models
\citep{Rodriguez-Puebla+17,Moster+18,Behroozi+19}. The consensus of these studies is that
$\fstar$ increases with mass to a peak, with $\fstar\sim 20\%$ at $\Mstar \sim 5 \times
10^{10}\Msun$ and $\Mh\sim 10^{12}\Msun$ (near the mass of the Milky Way), and then decreases
with mass.

The standard explanation for the inverted-U shape of the SHMR is that feedback by young stars
is responsible for the low-mass part, while feedback from active galactic nuclei (AGN) is
responsible for the high-mass part. Both types of feedback are potentially capable of driving
outflows from a galaxy and impeding further inflows, thus quenching star formation. The effect
of stellar feedback on the SHMR is fairly well understood: a higher fraction of gas is driven
out of low-mass galaxies because they have lower escape speeds \citep[e.g.][]{DekelSilk86,
Veilleux+05}.
Near the peak of the SHMR, much of the gas probably circulates in a self-regulated fountain,
without escaping from the halo (e.g. \citealt{Tumlinson+17}, and references therein). The
effect of AGN feedback on the SHMR is less well understood, but it is plausible that it
drives energetic outflows that heat some of the circumgalactic gas, thus slowing or reversing
its inflows \citep[e.g.][]{Fabian12,KingPounds15,Harrison17}.
Mergers may also disrupt the inflow and quench star formation, at least temporarily
\citep[e.g.][]{Hopkins+10}. Both mergers and AGN feedback may cooperate to cause the decline of
the SHMR at high mass \citep[e.g.][]{Croton+06} since mergers can funnel gas to a central black
hole, igniting AGN feedback \citep[e.g.][]{Hopkins+06}, while also building galactic spheroids
\citep[i.e. classical bulges,][]{Hopkins+10b}.

In practice, the SHMR is usually assumed to be independent of galactic morphology
\citep{WechslerTinker18}. This assumption, however, appears to contradict the reasoning above
about the different roles of stellar and AGN feedback and the observation that the masses of
central black-holes correlate with the bulge masses of their host galaxies \citep{KormendyHo12}.
Thus, if AGN feedback is important, it should have more effect on the high-mass shape of the SHMR
for bulge-dominated galaxies than it does for disc-dominated galaxies. More specifically, $\fstar$
should decline with $\Mstar$ and $\Mh$ past the peak in early-type galaxies but rise or level off
in late-type galaxies. The main goals of this paper are to confirm this expected dependence of the
SHMR on galaxy morphology and to explore some of its implications for our understanding of galaxy
formation.

There is already some evidence for secondary correlations between the SHMR and other properties
of galaxies. This evidence comes from weak lensing
\citep{Mandelbaum+06,Mandelbaum+16,Tinker+13,Hudson+15,Taylor+20}, satellite kinematics
\citep{Conroy+07,More+11,WojtakMamon13,Lange+19}, empirical models
\citep{Rodriguez-Puebla+15}, abundance matching \citep{Hearin+14,Saito+16} or a mix of these
\citep{Dutton+10}. The results of these studies are consistent with the expectation that
early-type galaxies occupy more massive halos than late-type galaxies of the same stellar mass.
However, in most cases, the results are based on stacking the observations in large samples of
galaxies to amplify the marginal or undetectable signals from individual galaxies, an approach
that can sometimes yield spurious correlations (and has led to some debate on the topic, see e.g.
S6.1 in \citealt{WechslerTinker18}). Some recent hydrodynamical simulations also display the
expected differences between the SHMR of early and late-type galaxies
\citep{Grand+19,Marasco+20,Correa+20}.

The most direct approach to deriving the SHMR is to estimate the masses of individual halos from
the observed kinematics of tracer objects whose space distribution extends well beyond the luminous
parts of galaxies. Since the available tracers almost never reach the expected outer (virial)
radii of the halos, estimates of their total masses require priors such as the correlation between
concentration and mass found in $\Lambda$CDM simulations. This is the approach used by
\citet[][hereafter \citetalias{PFM19}]{PFM19} to derive the SHMR of 110 late-type galaxies with
extended HI rotation curves in the Spitzer Photometry and Accurate Rotation Curves (SPARC) sample
\citep[][]{SPARC}\footnote{In this context, HI is a better tracer than H$\alpha$
because it usually extends to larger radii \citep{vanAlbada+85,Kent87}.}.
\citetalias{PFM19} found that the SHMR rises monotonically for all masses and reaches
$\fstar \sim 0.3$--1 for the most massive galaxies in this sample, with $\Mstar \sim 1$--$3 \times
10^{11} M_\odot$ (dubbed the ``failed feedback problem''). This result is in stark contrast to the
declining high-mass form of the SHMR found in most studies of the general population of galaxies,
which is dominated by early types at the highest masses \citep{Kelvin+14}.

In this paper, we derive the SHMR for early-type galaxies by methods as similar as possible to
those used by \citetalias{PFM19} for late-type galaxies. In particular, we adopt the same
$\Lambda$CDM correlation between halo concentration and mass. However, instead of using HI rotation
curves to probe the gravitational potential, we use the radial velocities of globular clusters (GCs)
around 25 massive early-type galaxies in the SAGES Legacy Unifying Globulars and GalaxieS Survey
\citep[SLUGGS,][]{Brodie+14}. We fit a distribution function, expressed in terms of action and angle
variables, to the observed kinematics and space distribution of each GC system to estimate its halo
mass. This enables us, for the first time, to make a direct and robust comparison between the SHMR
of early types and late types based on individual estimates of halo masses.

The different SHMRs for early-type and late-type galaxies -- one with a prominent bend, the other
without -- may seem puzzling because both galactic types have similar scaling relations between
stellar velocities and masses (the Faber-Jackson and Tully-Fisher relations). Since the SHMR and
the velocity scaling relations both depend on $\Mstar$, one might reasonably expect a bend in the
former to impose a bend in the latter. However, since the SHMR also depends on $\Mh$, it is possible
that the similar velocity scaling relations are actually explained by, and disguised by, different
underlying relations between the stellar velocities of early-type and late-type galaxies and one or
more properties of their dark matter halos, thus offering potentially important clues about the
physical mechanisms responsible for different galactic morphologies. We explore this issue here for
the first time.

The remainder of this paper is organised as follows. In Section~\ref{sec:methods}, we summarise the
data and dynamical models we use to estimate $\Mh$ for early types and the analogous estimates of
$\Mh$ from \citetalias{PFM19} for late types. Interested readers can find a full description of
our models for early-type galaxies in Appendix~\ref{app:model}.
In Section~\ref{sec:shmr}, we present our SHMR, showing unambiguously that it depends on galaxy
morphology and disc fraction. Section~\ref{sec:comp} compares our results with previous evidence
for different SHMRs. In Section~\ref{sec:SL}, we reconcile the different shapes of the SHMRs for
late-type and early-type galaxies with their similar velocity scaling relations in terms of
differences between their stellar and halo velocities, and we interpret this result as a natural
consequence of the different roles of smooth inflow, merging, and AGN feedback in the formation of
galactic discs and spheroids. Section~\ref{sec:concl} summarises our main conclusions.

Throughout the paper, we use a fixed critical overdensity parameter $\Delta=200$ to define
the virial quantities of dark matter halos and a standard $\Lambda$CDM model with a Hubble constant
$H_0=67.4$ km s$^{-1}$ Mpc$^{-1}$ \citep{Planck18}. We distinguish late-type and early-type galaxies
based on published morphological classifications: early types are E and S0 (Hubble type
$T<0$), while late types are S0/a, Sa, Sb, Sc, Irr (Hubble type $T\geq 0$).

\begin{figure*}
\begin{center}
\includegraphics[width=\textwidth]{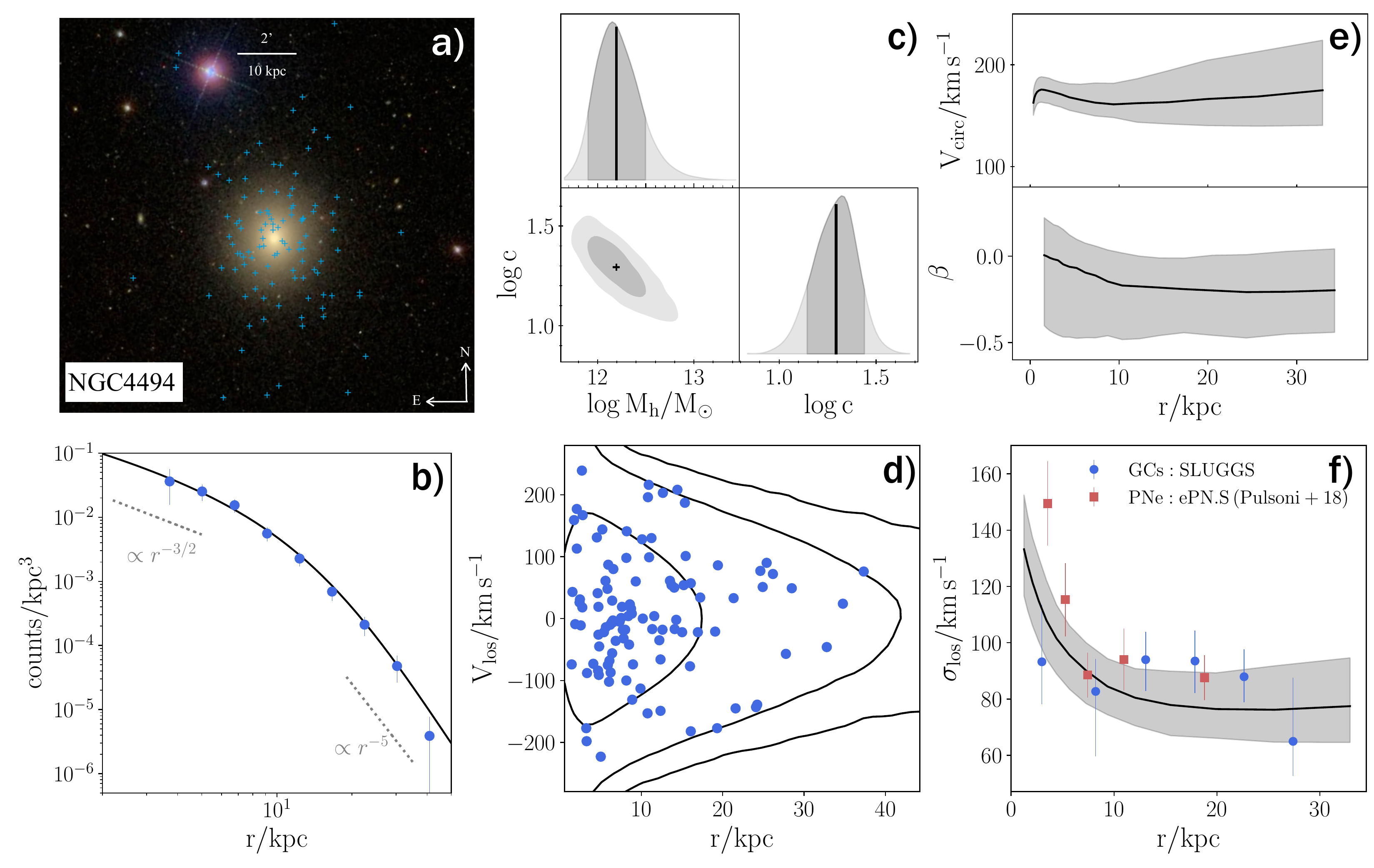}
\end{center}
\caption{Illustration of our modelling technique using NGC 4494 as an example. In all
         panels, GC data are shown as blue points, the DF model is shown as a black
         solid curve, and the 68\% confidence interval of the model is shown as a grey
         shaded area.
         {\bf a)} SDSS colour image of the galaxy with blue crosses marking the
         spectroscopically confirmed globular clusters by the SLUGGS survey
         (made with \texttt{Aladin}).
         {\bf b)} Projected number density profile of the GCs compared to that of the
         $f(\bJ)$ model.
         {\bf c)} Marginalised posterior distribution of the halo mass and concentration
         estimated with an MCMC method. Dark and light grey areas encompass respectively
         68\% and 95\% probability, while the black solid lines and cross mark the
         maximum-likelihood model.
         {\bf d)} Line-of-sight velocity as a function of projected radius, $\Vlos-r$, for
         the GCs compared with the projected phase-space density of the
         maximum-likelihood $f(\bJ)$ model. The contours contain 68-95-99\% of the projected
         phase-space density of the model.
         {\bf e)} Circular velocity profile, $V_{\rm circ}$ (top), and velocity anisotropy
         profile, $\beta=1-(\sigma_\theta^2+\sigma_\phi^2)/2\sigma_r^2$ (bottom), of the
         $f(\bJ)$ model.
         {\bf f)} Line-of-sight velocity dispersion profile of the GCs measured with SLUGGS
         data compared to that of our $f(\bJ)$ model.
         We also compare to independent measurements of the $\sigmalos$ profile of planetary
         nebulae obtained by \citet[][red squares]{Pulsoni+18}.
        }
\label{fig:showcase_ngc4494}
\end{figure*}

\begin{figure}
\begin{center}
\includegraphics[width=0.45\textwidth]{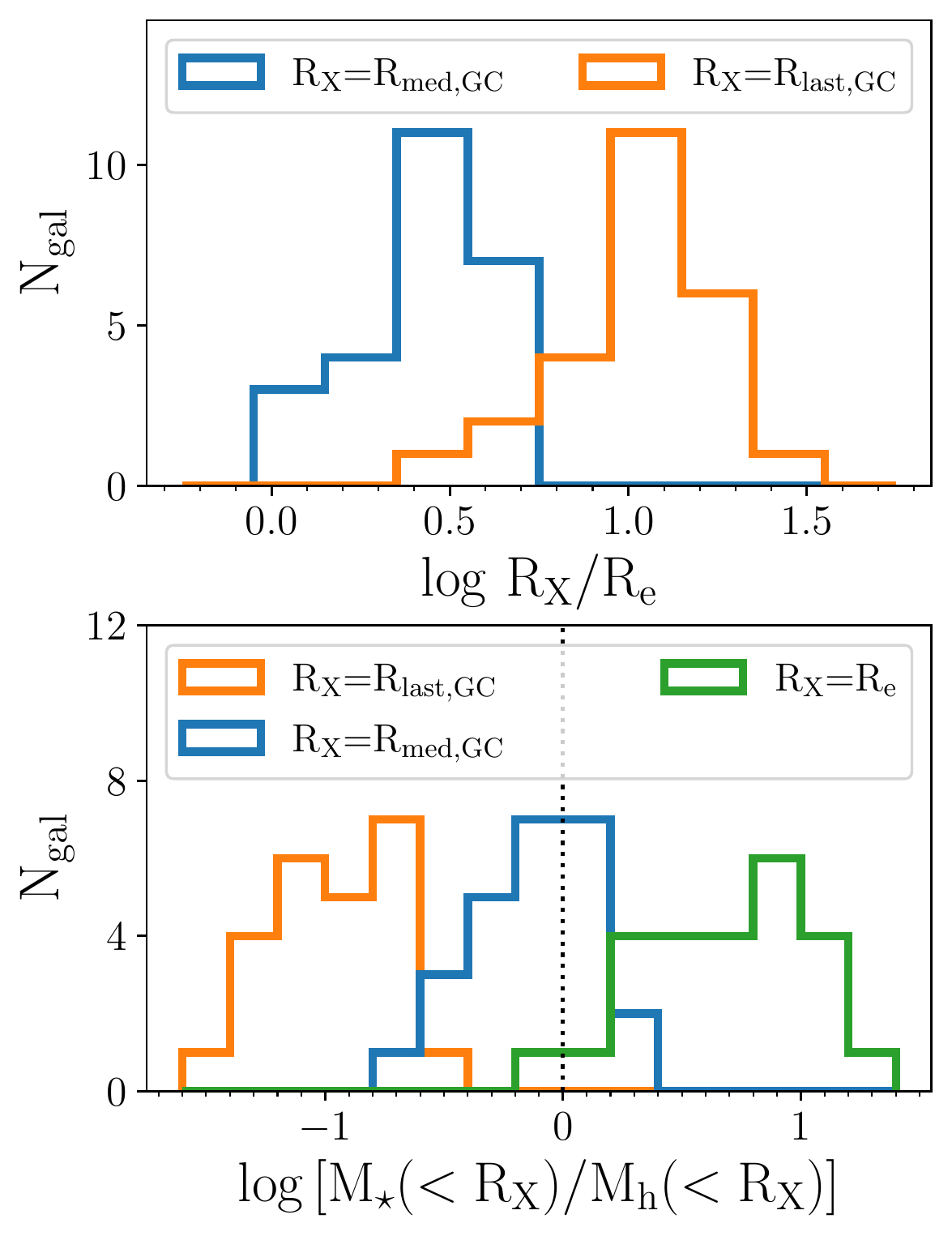}
\end{center}
\caption{\textit{Top panel}: distributions of the median radii
         ($R_{\rm med,GC}$, blue) and of the outermost radii ($R_{\rm last,GC}$,
         orange) of the GC systems around the early types in SLUGGS. Both radii are
         normalised to the effective radius of the luminous galaxy ($R_{\rm e}$).
         \textit{Bottom panel}: distributions of the ratios of stellar to halo
         mass enclosed within the effective radii ($R_{\rm e}$, green),
         within the median radius of the GC system ($R_{\rm med,GC}$, blue), and
         within the outermost GC radius ($R_{\rm last,GC}$, orange). These are
         computed from the best-fit $f(\bJ)$ models of each galaxy. The dotted
         vertical line separates the region where the dark matter dominates (left)
         and where the stars dominate (right).
        }
\label{fig:radii_MsMh}
\end{figure}

\section{Dynamical estimates of halo masses for early-type and late-type galaxies}
\label{sec:methods}

\subsection{Early types}

The main novelty of this work is our dynamical estimates of halo masses ($\Mh$) for
individual nearby ellipticals and lenticulars. To derive these, we use observations of the
kinematics of the GC systems around these galaxies and we model explicitly their distribution
function. Our method is an adaptation of that of \cite{PostiHelmi19}, who used it to measure
the halo mass of the Milky Way. The method consists of two main ingredients: the distribution
function (DF) of the GC system and the gravitational potential. Here we provide an overview of
our method with the guidance of Fig.~\ref{fig:showcase_ngc4494}, which illustrates
the input, fitting, and output of our model for a representative galaxy, NGC 4494. This section
introduces all the information needed to understand our results, while Appendix~\ref{app:model}
provides a detailed description of our model, which the busy reader can skip.

{\it Data.} We take the velocity and position data of the globular cluster systems around 27
nearby bright ellipticals and lenticulars from the SLUGGS Survey \citep{Brodie+14}. The
centrepiece of this data set is the catalogue of radial velocities of the
spectroscopically-confirmed GCs obtained with DEIMOS@Keck \citep[][]{Forbes+17b}. Each galaxy
has tens or hundreds of GCs, with a significant galaxy-to-galaxy variation and with a typical
uncertainty on each radial velocity of about $\sim 10-20$ km/s. The radial coverage of the GC
system is also quite varied: the radius containing 90\% of the observed GCs ranges between 8 and
98 kpc (corresponding to $3R_{\rm e}$ and $14R_{\rm e}$) with a median of about 30 kpc
(corresponding to $7.5 R_{\rm e}$).
The top panel of Fig.~\ref{fig:radii_MsMh} shows the distributions of the median radii
($R_{\rm med,GC}$) and the outermost radii ($R_{\rm last,GC}$) of the GC systems, both
normalised by the effective radii of the galaxies ($R_{\rm e}$).
To model the distribution of baryons, we use the photometric profiles derived from Spitzer
Space Telescope images at 3.6 $\mu$m by \cite{Forbes+17a}. Out of the 27 galaxies in
\cite{Brodie+14}, we exclude NGC 4474, since it does not have Spitzer images, and NGC 4111,
since it has fewer than 20 GC velocities.
As an example, in Fig.~\ref{fig:showcase_ngc4494}a, we show the distribution of the confirmed
SLUGGS GCs around NGC 4494.

\begin{figure*}
\begin{center}
\includegraphics[width=\textwidth]{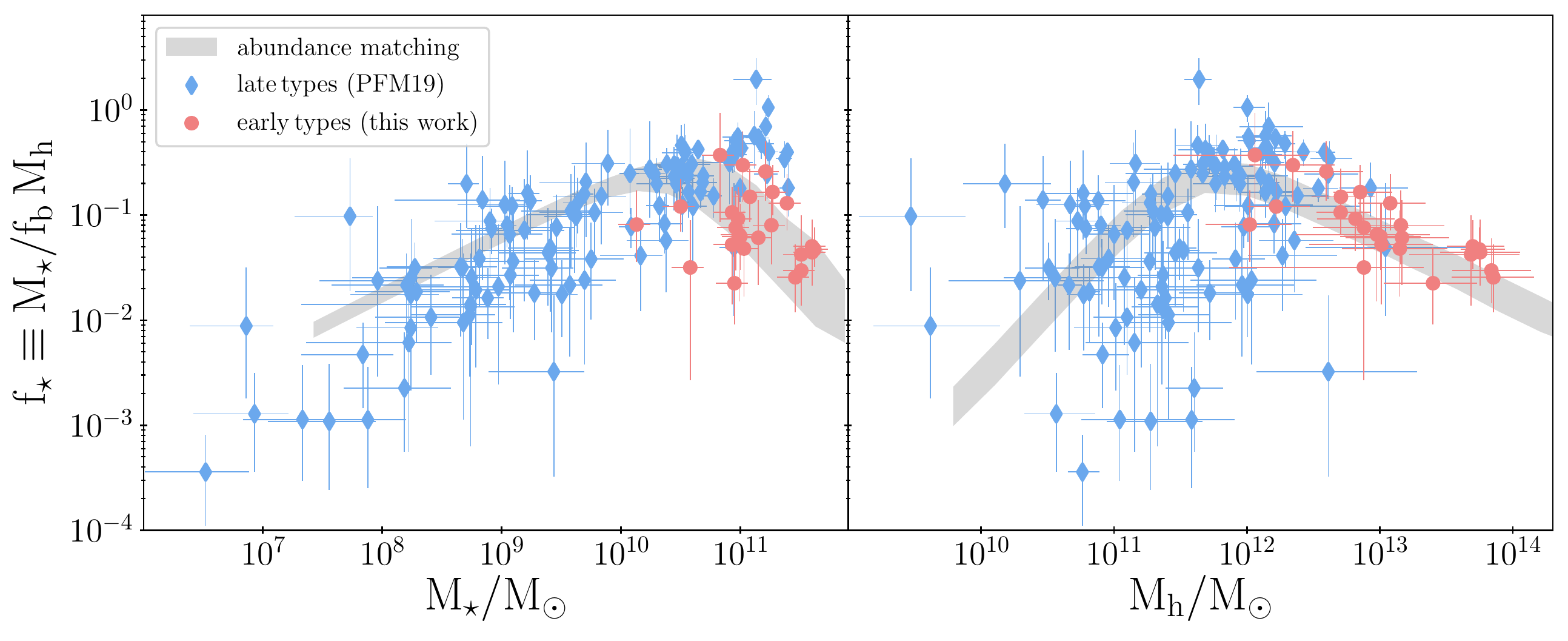}
\end{center}
\caption{SHMR in the form of the ratio $\fstar\equiv\Mstar/\fb\Mh$
         as a function of stellar mass (left) or halo mass (right) for the sample of spiral
         galaxies in SPARC (blue diamonds, \citetalias{PFM19}) and for the sample of ellipticals
         and lenticulars in SLUGGS (red circles, this work). The halo masses of late types are
         estimated from HI rotation curves, those of early types from the kinematics of the
         GC system. We compare to the SHMR from the abundance matching model by
         \citet[][grey band]{Moster+13}.
        }
\label{fig:fstar-SLUGGS-SPARC}
\end{figure*}

{\it Distribution function.} We use analytic DFs that depend on action-angle variables in the
form introduced by \cite{Posti+15} specifically to describe spheroidal systems. We refer to
these models as $f(\bJ)$, where $f$ is the DF and $\bJ$ are the action integrals.
The models we use have spherical space distributions but anisotropic velocity
distributions. The DFs are double power laws in the actions such that they generate
double power-law density distributions in physical space. Once the two slopes of the DF are
fitted to the space distributions of GCs (Fig.~\ref{fig:showcase_ngc4494}b), the three remaining
parameters specify the velocity distribution of the GC system.

{\it Gravitational potential.} The total potential in the $f(\bJ)$ model is a superposition of
two spherical components: the luminous galaxy and the dark matter halo. The galaxy is modelled
as a de-projected \cite{Sersic68} profile, based on the the photometry by \cite{Forbes+17b}, with
an adjustable mass-to-light ratio. The dark matter halo is assumed to have a
\citet[][herafter NFW]{NFW} profile parametrized by its virial mass ($\Mh$) and concentration
($c$). For the mass-to-light ratio, we impose a gaussian prior with a mean derived by
\cite{Forbes+17b}, using stellar population models, and a dispersion $\sigma_{\log (M/L)}=0.2$ dex.
For the halo concentration, we impose the correlation between $c$ and $\Mh$ found in
N-body $\Lambda$CDM simulations \citep{DuttonMaccio14}.

{\it Bayesian parameter estimation.} We use a Markov Chain Monte Carlo (MCMC) method to derive
posterior probabilities of the free parameters of our model (Fig.~\ref{fig:showcase_ngc4494}c).
The likelihood is given by the product of the DF convolved by the error distribution for each
cluster. Since we need 3 positions and 3 velocities to evaluate the DF, we sample the missing
position from the observed density distribution of the clusters and the two missing velocities
uniformly in the range allowed by the escape speed of the model
(Fig.~\ref{fig:showcase_ngc4494}d).
We then use our $f(\bJ)$ models to derive the intrinsic properties of the potential
(the circular velocity $V_{\rm circ}$) and of the GC system (the anisotropy parameter
$\beta=1-(\sigma_\theta^2+\sigma_\phi^2)/2\sigma_r^2$) shown in Fig.~\ref{fig:showcase_ngc4494}e.
In the bottom panel of Fig.~\ref{fig:radii_MsMh}, we use the $f(\bJ)$ models to compute the
ratios of stellar to halo mass enclosed within progressively larger radii: the luminous
$R_{\rm e}$, the median GC radius $R_{\rm med,GC}$, and the outermost GC radius
$R_{\rm last,GC}$.
Evidently, dark matter is negligible relative to stars near $R_{\rm e}$, is comparable near
$R_{\rm med,GC}$ ($\sim 2-3 R_{\rm e}$), and is dominant near $R_{\rm last,GC}$
($\sim 5-20 R_{\rm e}$).
We can also compute a posteriori the model line-of-sight velocity dispersion profile
and check that it is consistent with the observed profile for GCs in the SLUGGS Survey
and planetary nebulae in the ePN.S Survey \citep{Pulsoni+18}. Fig.~\ref{fig:showcase_ngc4494}f
shows this consistency for NGC4494.

\subsection{Late types}

\citetalias{PFM19} determined halo masses for a sample of nearby spirals by fitting galaxy
plus halo models to their extended HI rotation curves. Here we just summarise their analysis
and results and refer the reader to their paper for full details.

{\it Data.} The sample consists of 110 nearby spirals with 3.6 $\mu$m Spitzer images and
HI rotation curves drawn from the SPARC database compiled by \citet[][see also the original
references therein]{SPARC}.
The rotation curves, taken from various sources in the literature, were derived from
interferometric HI observations that extended well beyond the optical discs of the galaxies.
The sample spans a large range in stellar masses, from dwarfs ($\Mstar\sim 10^7 \Msun$) to
giants ($\Mstar\sim 10^{11} \Msun$).

{\it Rotation curve decomposition.} The observed rotation curves are decomposed into
gas, stars, and dark matter components. The gas contribution is derived directly from
the HI flux, while the stellar contribution is computed from the 3.6 $\mu$m photometry
with an adjustable mass-to-light ratio. The dark matter halo is modelled as a standard
NFW profile with variable virial mass ($\Mh$) and concentration ($c$) following the $c-\Mh$
relation from $\Lambda$CDM simulations \citep{DuttonMaccio14}.
In massive spirals, which are the focus of this work, models with cuspy dark matter
halos (such as NFW) match the observed rotation curves, yielding fits that are statistically
indistinguishable from those obtained with other halo models (e.g. pseudo-isothermal or cored
halos, see \citealt{deBlok+08}; \citealt{Martinsson+13}; \citealt{Katz+17}; \citetalias{PFM19};
\citealt{Li+20}).
This contrasts with the situation for dwarf galaxies, whose rotation curves are often
matched better with cored halo models \citep[e.g.][]{deBlok+01,Oh+11}.

{\it Bayesian parameter estimation.} The MCMC approach is used to fit the rotation curve and to
estimate the posterior distribution of the three free parameters of the gravitational potential:
the stellar mass-to-light ratio, the halo mass, and concentration.

It is important to note here that the key assumptions of the dynamical models for late types and
early types are the same: an adjustable mass-to-light ratio at 3.6 $\mu$m, a spherical NFW halo,
a prior following the $\Lambda$CDM $c-\Mh$ relation and an MCMC approach to sample the posterior.
This makes the results from our $f(\bJ)$ models for early-type galaxies directly comparable with
those of \citetalias{PFM19} for late-type galaxies.

\section{The SHMR for different galaxy types} \label{sec:shmr}

\subsection{Dependence on stellar and halo masses} \label{sec:ltg-etg}

In Fig.~\ref{fig:fstar-SLUGGS-SPARC}, we plot our estimates of $\fstar$
versus stellar mass $\Mstar$ (left panel) and halo mass $\Mh$ (right panel) for the 25
SLUGGS early-type galaxies. We compare these with the estimates of $\fstar$ from
\citetalias{PFM19} for SPARC late-type galaxies and with the abundance matching model
of \cite{Moster+13}.
We find that at a fixed stellar mass, above $\sim 5\times 10^{10} \Msun$, early types
have systematically lower $\fstar$ than late types of similar stellar mass, by a factor
of $\sim 7$ at $\Mstar \sim 10^{11}\Msun$.

\begin{figure}
\begin{center}
\includegraphics[width=0.45\textwidth]{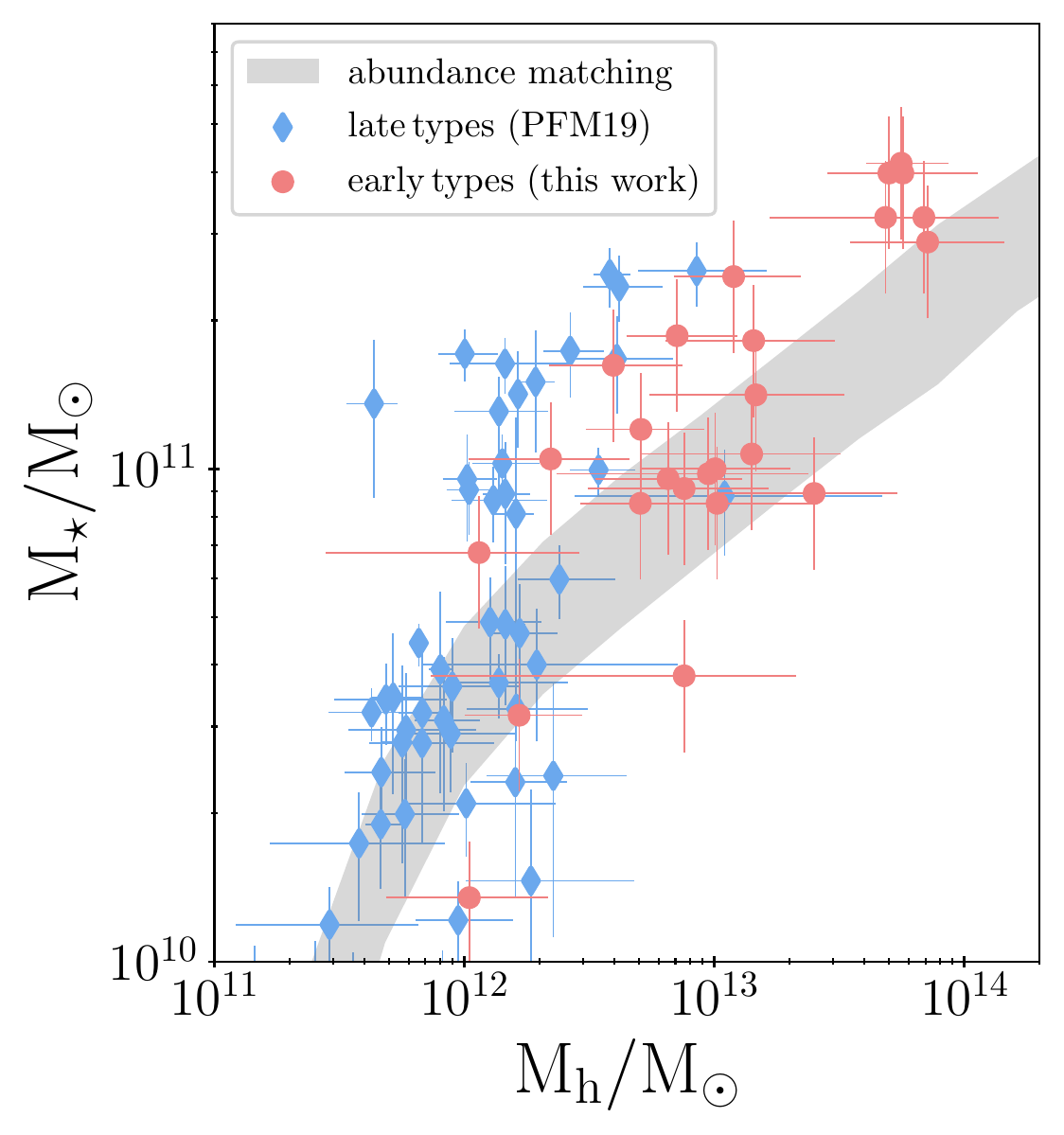}
\end{center}
\caption{SHMR in the form of stellar mass ($\Mstar$) as a function of halo mass ($\Mh$).
         Symbols are as in Fig.~\ref{fig:fstar-SLUGGS-SPARC}, however here we zoom
         in on the high-mass regime of the SHMR.
        }
\label{fig:Mstar-Mh-SLUGGS-SPARC}
\end{figure}

In order to guard against the possibility that the trends visible in
Fig.~\ref{fig:fstar-SLUGGS-SPARC} are induced by correlated errors in the plotted variables,
$\Mstar/\Mh$ versus $\Mstar$ or $\Mh$, we also show in Fig.~\ref{fig:Mstar-Mh-SLUGGS-SPARC}
the SHMR directly in the form $\Mstar$ versus $\Mh$. In particular, we zoom in on the
high-mass regime of the SHMR ($\Mstar>10^{10}\Msun$), which is of most interest here.
Fig.~\ref{fig:Mstar-Mh-SLUGGS-SPARC} confirms that late types and early types are separated
from each other in the same way as indicated in Fig.~\ref{fig:fstar-SLUGGS-SPARC}; massive
late types occupy systematically less massive halos than early types of the same stellar mass.

The grey bands in Fig.~\ref{fig:fstar-SLUGGS-SPARC}-\ref{fig:Mstar-Mh-SLUGGS-SPARC},
representing the conventional SHMR derived by abundance matching, are displayed only for
comparison purposes. We stress that all the main results of this paper come from dynamical
analyses of late-type and early-type galaxies and do not depend in any way on abundance
matching.
In these figures, we show the SHMR from \cite{Moster+13} because it represents a
consensus in the field \citep[see Fig. 2 in][]{WechslerTinker18}. In the SHMR derived by
\cite{Kravtsov+18}, massive galaxies tend to occupy slightly less massive
halos with respect to the \cite{Moster+13} SHMR. However, even in this case, the qualitative
picture presented in Fig.~\ref{fig:fstar-SLUGGS-SPARC}-\ref{fig:Mstar-Mh-SLUGGS-SPARC} remains
valid.

\begin{figure*}
\begin{center}
\includegraphics[width=\textwidth]{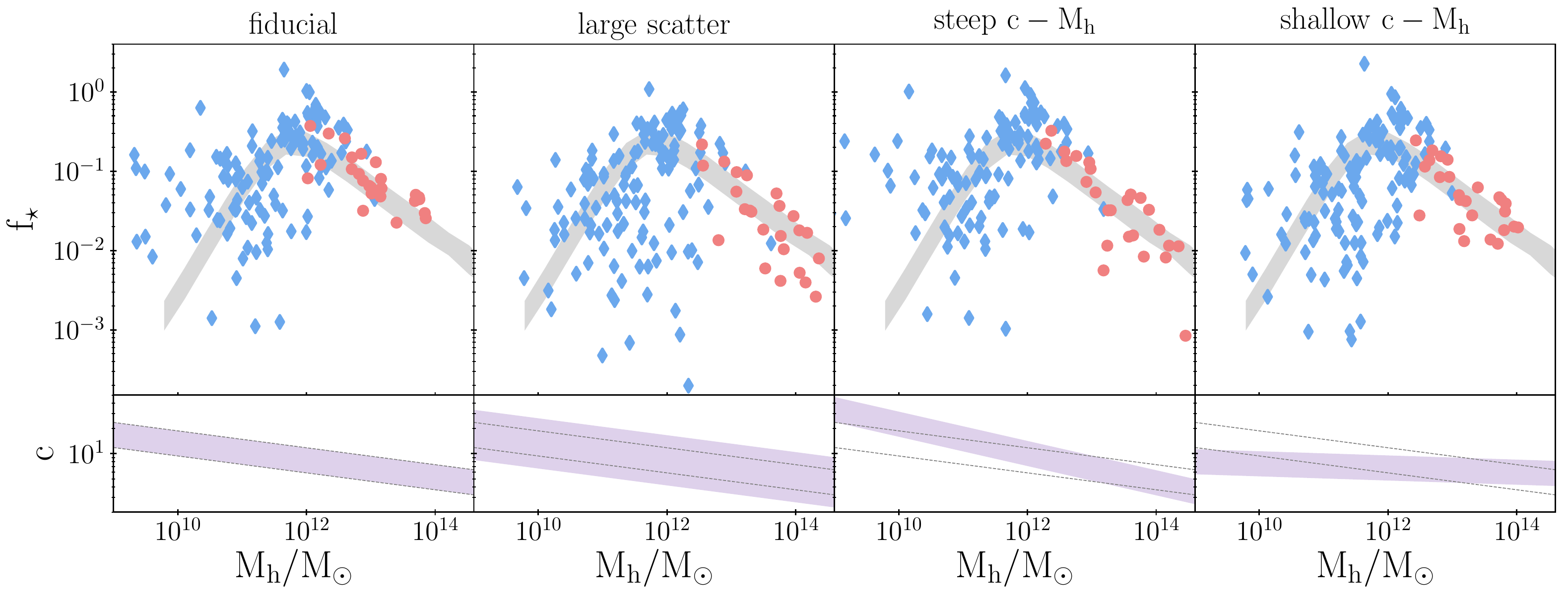}\\
\includegraphics[width=\textwidth]{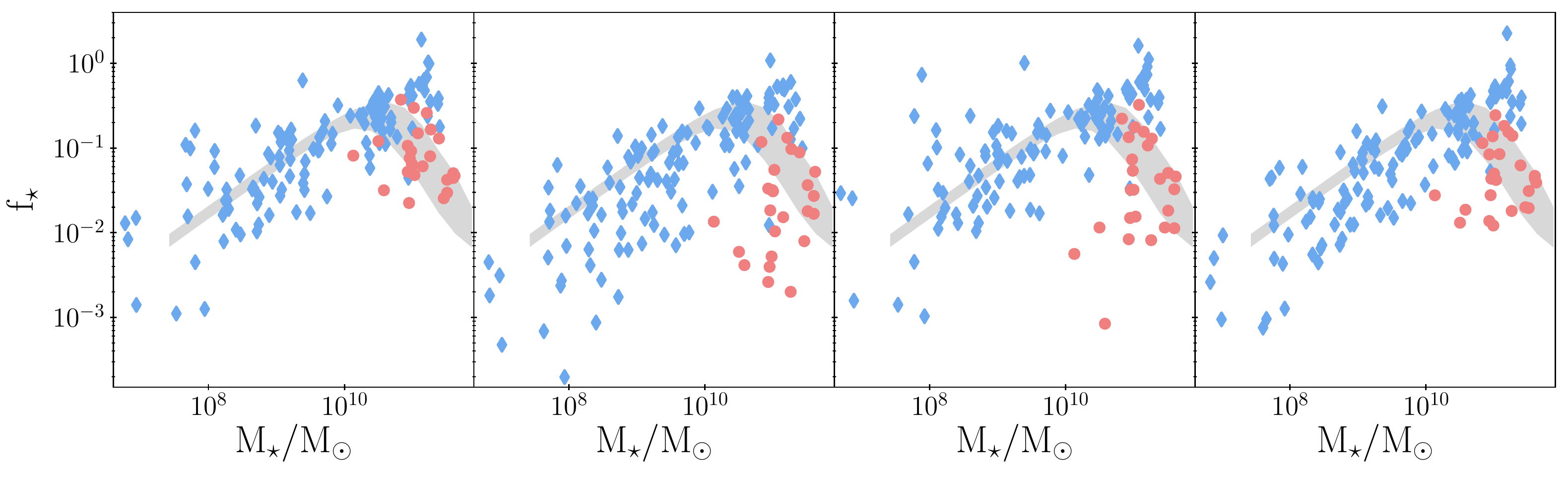}
\caption{Effect of varying the prior on the $c-\Mh$ correlation in our dynamical
         determination of the SHMR.
         The top (bottom) panels show the resulting $\fstar-\Mh$ ($\fstar-\Mstar$)
         relation, where the halo masses of both late types (blue diamonds) and early types
         (red circles) are calculated with a prior on the $c-\Mh$ relation that follows
         the purple shaded area ($1\sigma$) in the middle panel.
         From left to right, the four columns are for the following priors:
         i) the fiducial $c-\Mh$ from cosmological simulations by \citet[][with
         slope=$-0.101$ and scatter=0.11 dex]{DuttonMaccio14}, ii) the same $c-\Mh$,
         but with about twice the scatter (0.25 dex), iii) a steeper one (slope=$-0.2$), and
         iv) a shallower one (slope=$-0.05$). The black dashed lines in the middle panels
         show the fiducial $c-\Mh$ relation for comparison. In the top and bottom rows
         the grey shaded area shows the SHMR from \citet{Moster+13} based on abundance
         matching.
        }
\label{fig:c-Mh_prior}
\end{center}
\end{figure*}

For early types with $\Mstar\gtrsim 10^{11}\Msun$, we measure a scatter of $\approx 0.4$
dex in $\fstar$ at a fixed $\Mstar$. This scatter reflects a combination of several effects,
which we assume to be independent to first order: i) the observational errors in the GC
velocities and the uncertainty in the velocity dispersion due to sparse sampling, ii) the
uncertainty in the mass-to-light ratio, iii) the scatter in the $c-\Mh$ relation, and iv)
the intrinsic scatter in $\fstar$ at fixed $\Mstar$.
The first term varies substantially from galaxy to galaxy, as it is related to the
signal-to-noise of the GC spectra and to the number of GCs observed, but for a
typical galaxy this is of the order of 25\%, i.e. 0.1 dex. The second term is of the
order of 0.1-0.2 dex \citep{Forbes+17a}. The third term is an output of
cosmological simulations and is $\approx 0.11$ dex \citep{DuttonMaccio14}. The fourth term
can be estimated from the conventional SHMR. We generate a population of halos from a
standard halo mass function \citep{Tinker+08} and we assign an $\Mstar$ to each halo
following the \cite{Moster+13} SHMR. They estimate the scatter of $\fstar$ to be 0.15 dex
at a fixed halo mass $\Mh$, which corresponds to the grey bands in $\fstar$ versus $\Mstar$
and $\Mh$ in Fig.~\ref{fig:fstar-SLUGGS-SPARC}, and in $\Mstar$ versus $\Mh$ in
Fig.~\ref{fig:Mstar-Mh-SLUGGS-SPARC}.
The resulting scatter at a fixed stellar mass $\Mstar$ is about 0.08 dex below the turnover
at $\Mstar\sim 5\times 10^{10}\Msun$, but then increases substantially, reaching about 0.34
dex at $\Mstar\sim 10^{11}\Msun$\footnote{
This happens as a result of the combination of the SHMR with the steeply declining halo
mass function. Above the peak, low-mass halos that are high-$\fstar$ outliers of the
$\fstar-\Mh$ relation are about as common as high-mass halos with a typical $\fstar$.
This not only increases the scatter at a fixed $\Mstar$, but it also increases the
average $\fstar$ at a fixed $\Mstar$ with respect to that obtained by inverting
$\fstar(\Mh)$ \citep[see e.g.][]{Moster+20}.
}. Combining these four effects, we can nicely explain the observed scatter of 0.4 dex
in our estimates of $\fstar$.

\begin{figure*}
\begin{center}
\includegraphics[width=0.99\textwidth]{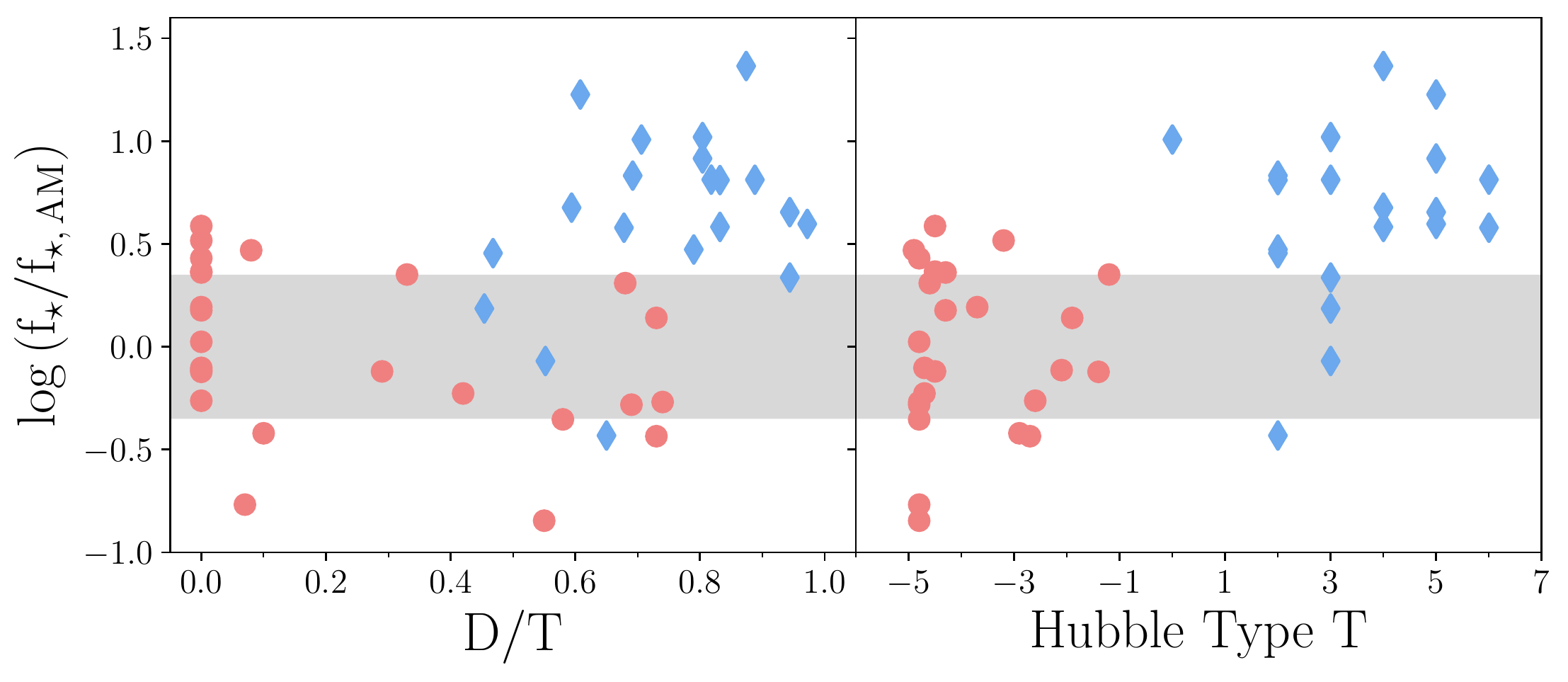}
\caption{Residuals of our dynamical estimate of $\fstar$ relative to the abundance-matching
         value $f_{\star,\,\rm AM}$ at the same stellar mass from \cite{Moster+13}
         versus the disc-to-total ratio $D/T$ (left) and Hubble type $T$ (right), for
         massive late types ($\Mstar>5\times 10^{10}\Msun$, blue diamonds)
         and early types (red circles).
         The grey area shows the scatter of the \cite{Moster+13} SHMR around
         $\Mstar\sim 10^{11}\Msun$. Several early types pile up at $D/T$=0 since no disc
         component could be clearly identified from their photometry.
        }
\label{fig:Deltafstar-DT}
\end{center}
\end{figure*}

Plotting $\fstar$ as a function of $\Mh$ demonstrates clearly that early types occupy halos
of a wide range of masses ($10^{12}\Msun\lesssim\,\Mh\lesssim 10^{14}\Msun$). In contrast,
late types of similar stellar mass are all found in halos of mass $\Mh \sim 10^{12}\Msun$ and
virtually none occupies halos more massive than $\Mh \sim 5\times 10^{12}\Msun$. This is
potentially a very important finding since it hints at the existence of an upper limit to the
masses of halos within which discs can form \citep[e.g.][]{DekelBirnboim06}.

Of all the assumptions in our modelling technique, we found that the prior on the $c-\Mh$
correlation has the largest effect on estimates of $\Mh$. In Fig.~\ref{fig:c-Mh_prior}, we
show the results of some tests to assess the robustness of our findings. We re-fitted our
$f(\bJ)$ models to the SLUGGS data with different priors for the halo concentration-mass
relation: in particular, we doubled the scatter and we increased and decreased the slopes
of the $c-\Mh$ relation so as to span the $1\sigma$ range of the the standard
$\Lambda$CDM relation over the range of halo masses probed here\footnote{
We also repeated this test with systematically larger and smaller concentrations, i.e. with
$c$ following the fiducial $c-\Mh$ relation $\pm 1\sigma$, finding similar
results to those shown in Fig.~\ref{fig:c-Mh_prior}}.
These priors are shown in the middle row of panels in Fig.~\ref{fig:c-Mh_prior}. Each column
of Fig.~\ref{fig:c-Mh_prior} shows the $\fstar-\Mh$ (top) and $\fstar-\Mstar$ (bottom)
relation that we obtain when assuming these different priors. To compare early types (red)
and late types (blue) consistently, we re-computed the rotation curve decompositions of
the late types in SPARC with each prior. From the results plotted in Fig.~\ref{fig:c-Mh_prior},
we note that, while there can be significant differences for individual galaxies, the
general trends for the populations of late types and early types remains robust.

These tests give us confidence that the systematic difference of the SHMR
of massive late-type and early-type galaxies is real. An important point that we need to
emphasise here is that the $c-\Mh$ priors that we have used for these tests are deliberately
extreme. In fact, such a large scatter (0.25 dex) or such steep or shallow slopes are outside
of the range of published $c-\Mh$ relations for the standard $\Lambda$CDM cosmogony
\citep[e.g.][]{DiemerKravtsov15}. This allows us to exclude, with high confidence, that the two
branches of the SHMR revealed in Fig.~\ref{fig:fstar-SLUGGS-SPARC} are the result of
late types and early types occupying halos with systematically different concentrations.

\subsection{Dependence on disc fraction and morphological type}

In Sect.~\ref{sec:ltg-etg}, we investigated the difference in $\fstar$ between two broadly
defined galaxy samples, late types and early types, while here we look in more detail at how
$\fstar$ depends continuously on disc fraction and morphological type for massive galaxies.

For 21 of the 25 early types in SLUGGS, we rely on the photometric bulge/disc
decompositions in the $r$-band performed by \cite{Krajnovic+13}, who fitted a S\'{e}rsic plus
exponential functions to the observed 1D photometric profiles. We note that, for half of these,
\cite{Krajnovic+13} found no significant contribution from an exponential disc component.
For the four remaining galaxies, we fitted the 1D $R$-band profiles from the Carnegie-Irvine
Galaxy Survey \citep{Li+11} with S\'{e}rsic plus exponential functions, in order to
perform a similar analysis to that of \cite{Krajnovic+13}.

\begin{figure*}
\begin{center}
\includegraphics[width=\textwidth]{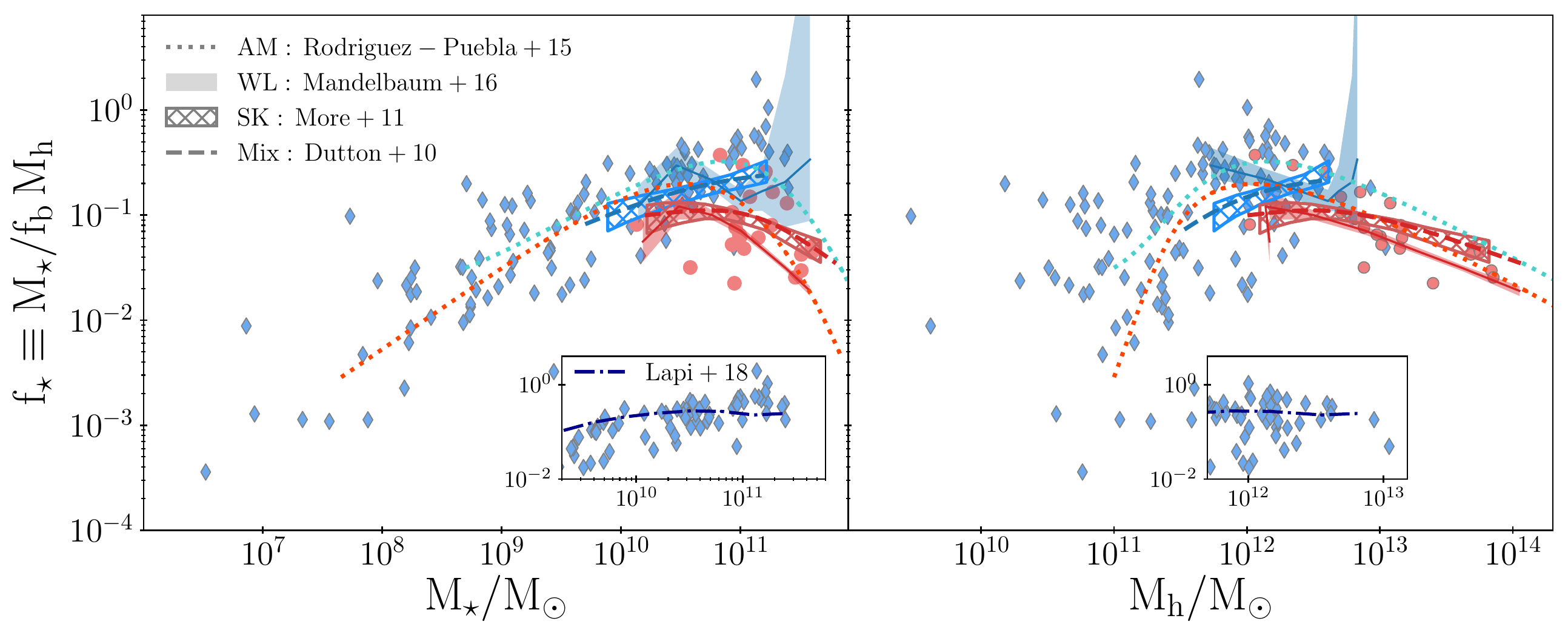}
\end{center}
\caption{Same as Fig.~\ref{fig:fstar-SLUGGS-SPARC}, but we compare our
         SHMR with others that are representative of studies based on different
         techniques: abundance matching \citep{Rodriguez-Puebla+15}, weak-lensing
         \citep{Mandelbaum+16}, satellite kinematics \citep{More+11} and a combination of
         the above \citep{Dutton+10}. Blue colour is used for late types, red colour is used
         for early types. The insets show the comparison to the estimates of $\fstar$ for
         late-type galaxies by \cite{Lapi+18}. We omit the errorbars for clarity.
        }
\label{fig:lit_comp}
\end{figure*}

We include only the 20 late types with the largest stellar mass for this comparison
($\Mstar > 5\times 10^{10}\Msun$). We take the disc-to-total ratios $D/T$ from either:
i) the Spitzer Survey of Stellar Structure in Galaxies \citep[][]{Sheth+10}, based on 2D
bulge/disc decompositions of the 3.6 $\mu$m Spitzer images with the code \texttt{galfit}
\citep{Peng+02,Peng+10} when available; ii) otherwise, from the kinematic decompositions
reported by \citet[][and references therein]{FR13,FR18}. For a few galaxies that do not
have $D/T$ from i) or ii), we used the decomposition of the 1D surface
brightness profiles at 3.6 $\mu$m performed by \cite{SPARC}.

For all of these massive galaxies, we calculate the residual of our estimate of $\fstar$
relative to the abundance-matching value $f_{\star,\rm AM}$ at the same mass from
\cite{Moster+13} and express it in the form $\log(\fstar/f_{\star,\,\rm AM})$.
We plot this quantity in Fig.~\ref{fig:Deltafstar-DT} as a function of $D/T$
(left) and as a function of the Hubble type $T$ (right), both for late types (blue
diamonds) and early types (red circles). We find that bulge-dominated galaxies ($D/T<0.2$,
$T<-3$) have small residuals with respect to abundance matching, as they lie within its scatter
(grey area). In contrast, disc-dominated galaxies ($D/T>0.8$, $T>4$) are found to have
systematically larger $\fstar$. The transition between these two regimes occurs at around
$D/T\sim 0.6$ or $T\sim 2$. The spirals with lowest disc fractions in this sample
($D/T\sim 0.5$, $T \sim 2$) are indeed in better agreement with abundance matching, although
the scatter of the points is substantial. The dependence of $\fstar$ on $D/T$ and on $T$ that
we observe in Fig.~\ref{fig:Deltafstar-DT} indicates that the location of a galaxy in the
$\fstar-\Mstar$ diagram depends fairly continuously on its disc fraction.

From this new perspective, we can now see that derivations of the SHMR that include galaxies
of all types are likely to overestimate the scatter at the high-mass end. This is because
the trend of $\fstar$ with $D/T$ (Fig.~\ref{fig:Deltafstar-DT}), if not recognised,  will
simply be counted as scatter. However, this effect is likely to be small since disc-dominated
galaxies are relatively rare at the high-mass end ($\sim 10\%$ at $\Mstar\gtrsim 10^{11}\Msun$,
see e.g. \citealt{Kelvin+14}; \citealt{Ogle+19a}).
To gain some intuition on the magnitude of this effect, we performed a simple calculation in an
extreme case of a binary population of galaxies: pure discs and pure spheroids. At a fixed
$\Mstar=10^{11}\Msun$, we have 90\% spheroids on the abundance matching relation, i.e. with
$f_{\star,\,\rm AM}(\Mstar)$, and we have 10\% discs with $\fstar$ systematically offset from
this by a factor 0.8 dex.
In order to match the scatter of the \cite{Moster+13} SHMR at that stellar mass, which is
$\approx$0.35 dex, we need to decrease the intrinsic scatter of the pure spheroid population
to $\approx 0.27$ dex. This exercise suggests that calculations that ignore the dependence of
the SHMR on morphology will overestimate the intrinsic scatter by about $0.08$ dex.

\section{Comparison with other estimates of the SHMR} \label{sec:comp}

In this section, we compare our derivation of the $\fstar-\Mstar$ relation for individual
late-type and early-type galaxies with other estimates from the literature.
Taken together, these results provide additional evidence that galaxies of different types
occupy halos of different masses.

\subsection{Statistical estimates of the SHMR}

In Fig.~\ref{fig:lit_comp}, we show how our SHMR compares to those derived by different
techniques, for late (blue) and early types (red) separately. We notice that there is
general agreement among these studies, with our SHMR showing perhaps the largest differences
between late and early types.

\begin{figure*}
\begin{center}
\includegraphics[width=\textwidth]{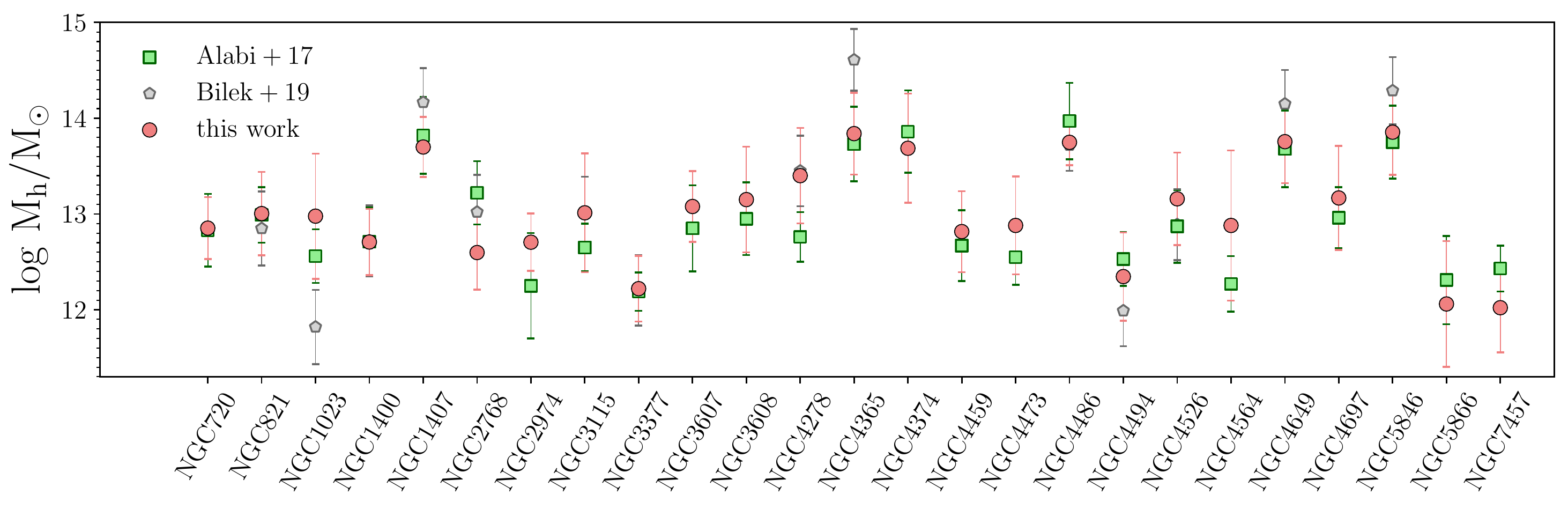}
\end{center}
\caption{Comparison of our estimates of halo masses for early-type galaxies in SLUGGS
         based on $f(\bJ)$ models (red circles) with those of \citet{Alabi+17}, using
         the TME (green squares), and those of \citet{Bilek+19}, based on a Jeans analysis
         (grey pentagons).
        }
\label{fig:a17_b19_comp}
\end{figure*}

Satellite kinematics \citep[e.g.][]{Conroy+07,More+11,WojtakMamon13,Lange+19} and especially
weak lensing \citep[e.g.][]{Mandelbaum+06,Mandelbaum+16,Tinker+13,Hudson+15,Taylor+20} are,
in principle, reliable tracers of halo masses out to very large radii. However, these methods
rely heavily on stacking (hundreds or thousands of) galaxies that are usually grouped into
late types and early types via a hard cut in colour. These analyses have the advantage of
including large numbers of galaxies but the disadvantage that colour is an imperfect proxy
for morphology, since it depends on a combination of other factors, including star formation
rate and history, dust reddening, and metallicity. Hence, the differences between
late types and early types will be artificially attenuated.
Similar considerations apply also to SHMRs based on empirical models, which are mostly
constrained by observed stellar mass functions, with only indirect estimates of halo masses
\citep[e.g.][]{Dutton+10,Rodriguez-Puebla+15,Behroozi+19,Moster+20},

Our work reinforces these results since it is based on careful estimates of the halo masses of
individual galaxies in samples of late types and early types specifically selected for dynamical
studies. Our derivation of the SHMR has opposite strengths and weaknesses with respect to the
statistical estimates above. It is therefore not surprising that we find a somewhat larger
difference in $\fstar$ between late-type and early-type galaxies.

\subsection{Individual estimates of halo masses}

\subsubsection{Early types}

The SLUGGS globular cluster data have already been used by \cite{Alabi+16,Alabi+17}
to estimate the total halo masses of the galaxies by a simpler method: the so-called tracer
mass estimator \citep[TME,][with original formulation by \citealt{BahcallTremaine+81}]
{Watkins+10}. In this case, the total mass is taken to be $M_{\rm TME} = C\,\hat{R}\hat{V}^2/G$
where $G$ is the gravitational constant, $\hat{R}$ and $\hat{V}$ are a characteristic radius
and velocity of the system and $C$ is a dimensionless constant of order unity that is
calibrated a priori with some simple assumptions.
We compare in Fig.~\ref{fig:a17_b19_comp} our halo masses with those of \cite{Alabi+17} on
a galaxy-by-galaxy basis, finding overall consistency within a factor of a few.

One of the shortcomings of the TME method is that it does not clearly partition between luminous
and dark components. This is of particular importance in the context of the SLUGGS early
types, since the TME is sensitive to the dynamical mass near the median radius of the tracer
population ($R_{\rm med,GC}$), which happens to be where the masses of stars and dark matter
are comparable (see Fig.~\ref{fig:radii_MsMh}). Therefore, it is not surprising that the $\Mh$
estimates of this method are consistent within a factor of a few with those of our $f(\bJ)$
models.

\cite{Bilek+19} recently estimated the halo masses of the early-type galaxies in SLUGGS,
by analysing the kinematics of their GC systems. In particular, they used the GC velocity
dispersion profiles together with the Jeans equations to constrain the gravitational potential.
They modelled the dark matter halo with a spherical NFW profile and they imposed a prior on the
$c-\Mh$ relation from cosmological simulations, as in our approach. However, rather than allowing
the data to constrain the velocity anisotropy of the GC system, as we do here, they assume a
$\beta$ profile as an input to their model.
We compare their estimates of halo masses with ours in Fig.~\ref{fig:a17_b19_comp}.
Despite the differences and limitations of their approach, we find that their results are
consistent with ours within the uncertainties. This comparison adds to our confidence that
our estimates of $\Mh$ are not biased by model-dependent systematic effects.


\subsubsection{Late types}

Several other studies have derived the SHMR of spiral galaxies from their rotation
curves, with results consistent with those from \citetalias{PFM19} shown here in
Fig.~\ref{fig:fstar-SLUGGS-SPARC}. \cite{Lapi+18} found a similar trend from stacked
H$\alpha$ rotation curves in an independent sample of spirals (see
Fig.~\ref{fig:lit_comp}). However, H$\alpha$ rotation curves typically do not extend
as far as HI rotation curves and thus, at best, provide only weak constraints on dark
matter halo masses \citep{vanAlbada+85,Kent87}.

\cite{Katz+17} also estimated the halo masses of spiral galaxies in the SPARC sample
from HI rotation curve decomposition. However, they either used an unconstrained fit
(i.e. with no prior on the $c-\Mh$ relation), leaving $\Mh$ undetermined, or they
imposed a prior on $\Mh$ from the \cite{Moster+13} SHMR. While the SHMR they derive
naturally follows closely the Moster et al. SHMR by construction, their estimates of
$\Mh$ are nevertheless in fair agreement with the ones we obtain without imposing a
prior on the SHMR.
\cite{Li+20} recently revised and expanded their approach to several different types
of halo profiles and reached similar conclusions.

It is also interesting to notice that a systematic difference between late types and
early types was also reported by \cite{Tortora+19} when looking at the dark matter
fraction $f_{\rm DM}$ within one effective radius $R_{\rm e}$ as a function of stellar
mass.
They noticed that late types lie on a decreasing $f_{\rm DM}-\Mstar$ relation, while
for massive early types this relation inverts, analogous to our results for $1/\fstar$
versus $\Mstar$ based on the total masses within the virial radii of the halos.
This suggests that, at a fixed $\Mstar$, massive discs are less dominated by dark
matter than spheroids, both globally and locally \citep[see also][]{Marasco+20}.

\subsection{The mass of the globular cluster system as a dark matter halo tracer}

Another novel method for estimating dark matter halo masses of galaxies is based on the
total mass of their GC system, $\MGCS$. Several recent studies have demonstrated the
existence of a convincing linear relation between $\MGCS$ and $\Mh$ for both
late types and early types \citep[e.g.][]{Blakeslee+97,SpitlerForbes09,Georgiev+10,
Harris+13,Harris+17,Hudson+14,BurkertForbes20}. The physical origin of this relation is
not well understood \citep[e.g.][]{KravtsovGnedin05,Boylan-Kolchin17,El-Badry+19}, but we
can exploit it as a semi-independent method for deriving the SHMR nonetheless\footnote{
A caveat we note here is that studies that determined the $\MGCS-\Mh$ relation often
assumed halo masses from a standard SHMR, typically not taking into account the morphology
dependence that we highlight in Fig.~\ref{fig:fstar-SLUGGS-SPARC}.
}.
Thus, we may regard the relation between $\LK/\MGCS$ and $\LK$ as a direct analogue
of the $\fstar-\Mstar$ relation, where $\LK$ is the $K$-band luminosity.
The $\LK/\MGCS-\LK$ relation requires only photometry, rather than spectroscopy of
the GCs, making it relatively easy to derive.
\citet[][see also references therein]{Harris+13} have assembled a
catalogue of $\MGCS$ for 422 nearby galaxies of all morphological types covering
a wide range in luminosities.

\begin{figure}
\begin{center}
\includegraphics[width=0.49\textwidth]{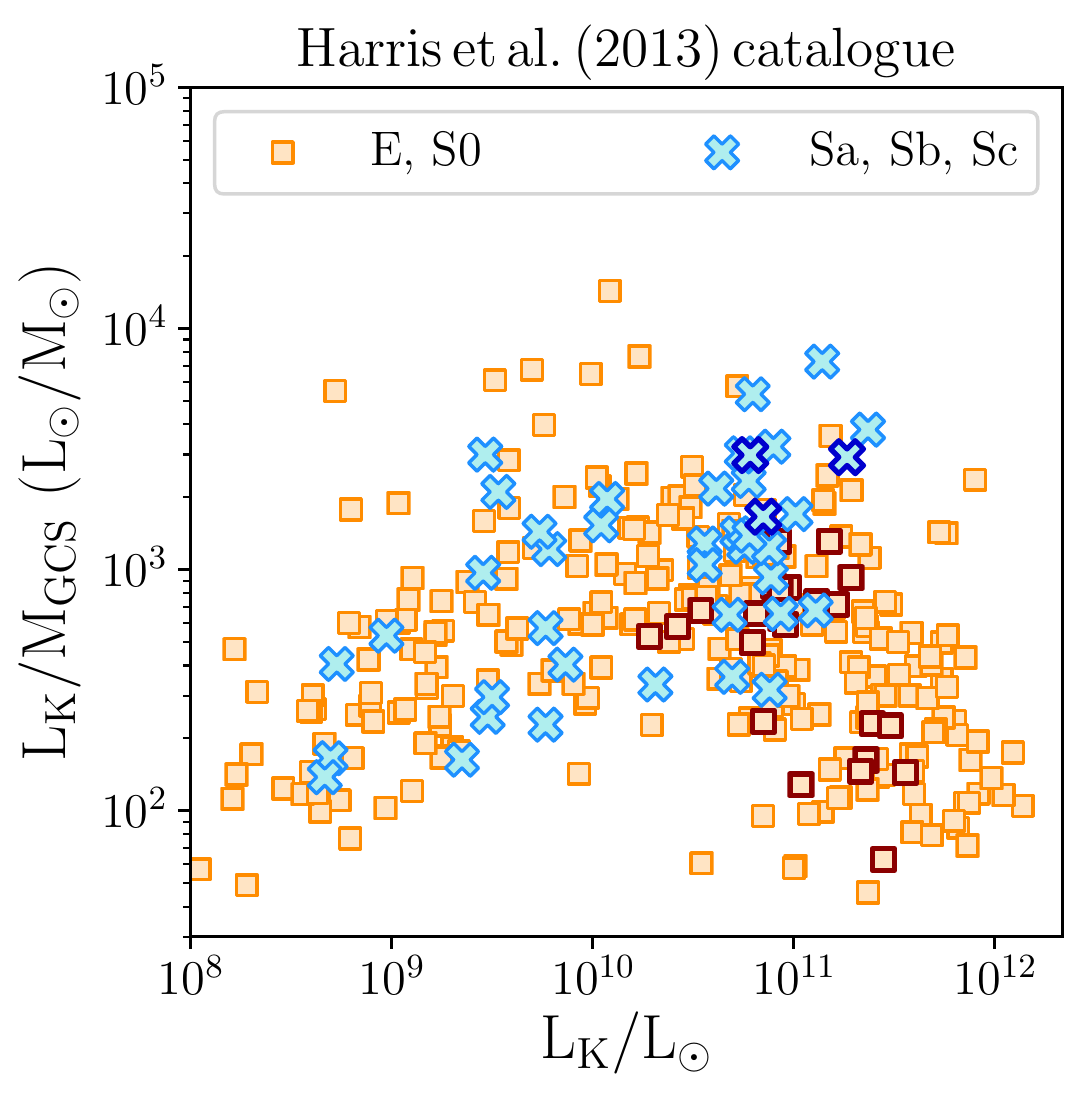}
\caption{Ratio of $K$-band luminosity to the mass of the GC system ($\LK/\MGCS$)
         versus $K$-band luminosity for galaxies in the \cite{Harris+13} catalogue. We plot
         spirals with turquoise crosses, and ellipticals and lenticulars with orange squares.
         We highlight the galaxies in common between the \cite{Harris+13} catalogue and our
         work with darker edge colours.
        }
\label{fig:Mgcs_H13}
\end{center}
\end{figure}

We plot in Fig.~\ref{fig:Mgcs_H13} the ratio $\LK/\MGCS$ as a function of $\LK$
for the galaxies in the \cite{Harris+13} sample, separating them into late types
(blue crosses) and early types\footnote{
Irregulars are excluded from this analysis
} (orange squares). The similarity between the $\LK/\MGCS-\LK$ relation in
Fig.~\ref{fig:Mgcs_H13} and the $\fstar-\Mstar$ relation in
Fig.~\ref{fig:fstar-SLUGGS-SPARC} is striking; early types turnover at about $\LK\sim
5\times 10^{10}\Lsun$, while bright late types seem to lie on a separate rising branch
\citep[see also][for a similar analysis of early types]{Kim+19}.

The \cite{Harris+13} sample contains 24 out of the 25 SLUGGS early types that we
analysed in this work, but only three of the 20 massive spirals in SPARC (NGC 891,
NGC 5907 and NGC 7331). The galaxies in common with our detailed analysis are indicated
in Fig.~\ref{fig:Mgcs_H13} by symbols with darker and thicker edges. We notice that the
three spirals have systematically higher $\LK/\MGCS$ than any of the early types of
similar $\LK$, in qualitative agreement with what we find on the $\fstar-\Mstar$ diagram.
From this analysis we conclude that, despite the large scatter, the $\LK/\MGCS-\LK$
relation is quite consistent with our more robust derivation of the $\fstar-\Mstar$
relation from GC kinematics and HI rotation curves.

\begin{figure*}
\begin{center}
\includegraphics[width=\textwidth]{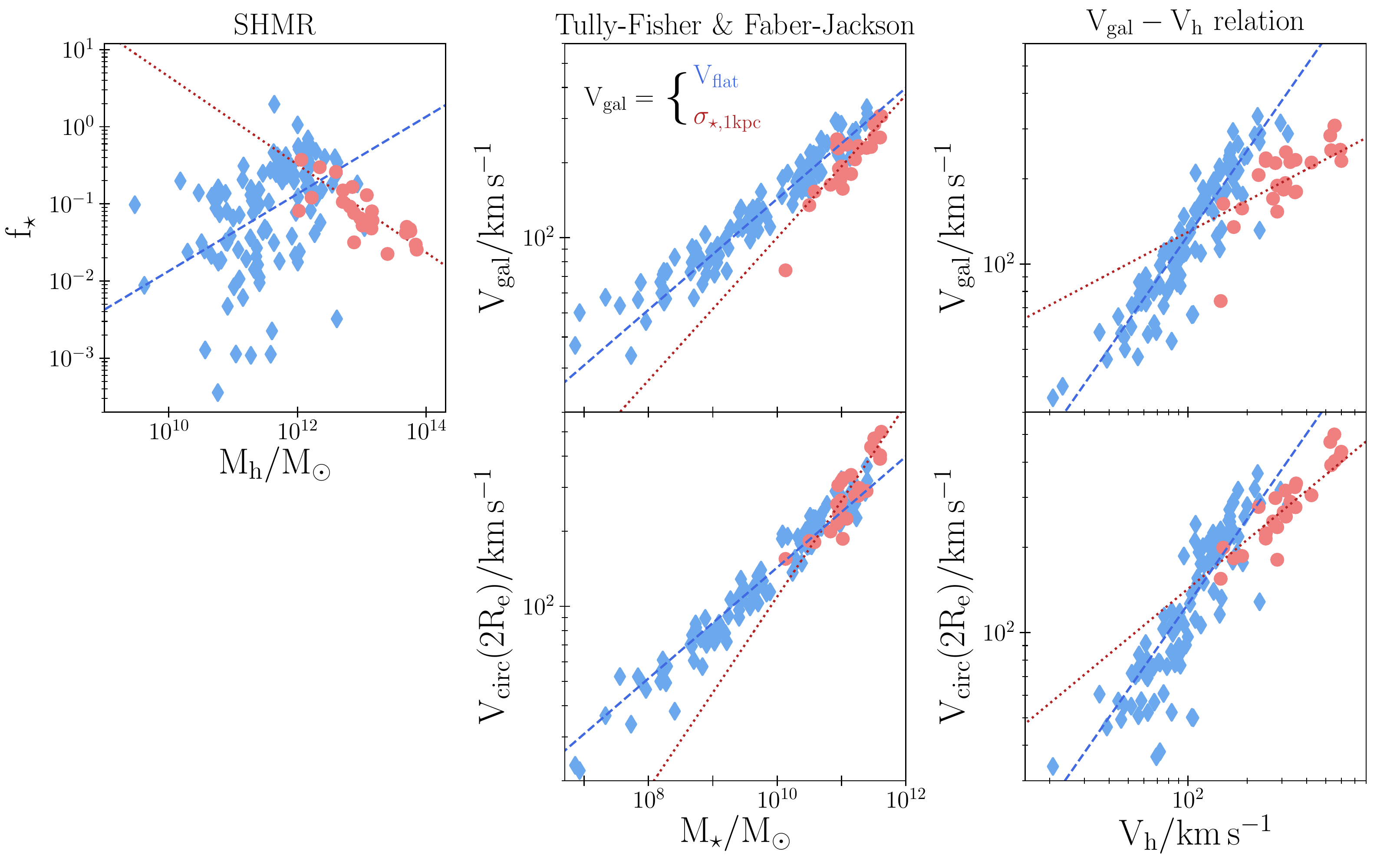}
\caption{SHMR ($\fstar-\Mh$, left), stellar mass-velocity scaling law ($\Mstar-\Vgal$, middle)
         and relation between galaxy velocity and halo velocity ($\Vgal-\Vh$, right) for
         the population of late types (blue) and early types (red). In the top-middle and
         top-right panels, the characteristic velocity of galaxies $\Vgal$ is the velocity
         along the flat parts of HI rotation curves for late types ($\Vflat$, \citealt{Lelli+16a})
         and the velocity dispersion of stars within 1 kpc for early types ($\sigFJ$,
         \citealt{Brodie+14}).
         In the bottom-middle and bottom-right panels, we substitute the observed $\Vgal$
         with the circular velocity evaluated at the same radius ($2R_{\rm e}$) for both
         galaxy types. In all panels, we show power-law fits to the data of late types (blue
         dashed lines) and early types (red dotted lines).
        }
\label{fig:TF-FJ-fstar}
\end{center}
\end{figure*}

\section{The SHMR and galaxy scaling laws} \label{sec:SL}

In the previous sections, we established that the SHMR of late types and early types
follows two distinct branches: one where $\fstar$ increases with mass for spirals, and another
one for ellipticals and lenticulars where $\fstar$ decreases beyond a peak near $\Mstar
\sim 5\times 10^{10}\Msun$.
Yet, late-type and early-type galaxies are known to obey very similar scaling relations between
their stellar masses and velocities; the \cite{TullyFisher77} and \cite{FaberJackson76}
relations are observed to be pure power laws with no significant features. These two facts
may appear to be at odds with each other, since velocity is a proxy for dynamical mass and
one might therefore expect the shape of the SHMR to impact the mass-velocity scaling laws
\citep[e.g.][]{Ferrero+17,Posti+19b}. How can we reconcile these seemingly contradictory
facts?

The resolution of this apparent paradox logically must involve the relations between the
characteristic velocities of galaxies and those of their dark matter halos, which need to
be different for late types and early types on the two branches of the SHMR.
In this section, we demonstrate this difference and discuss its physical implications.

\subsection{A paradox and its resolution} \label{sec:paradox}

We take measurements of the flat parts of HI rotation curves for spirals in SPARC
\citep[$\Vflat$, ][]{Lelli+16a} and stellar velocity dispersions measured within a fixed
radius of 1 kpc for the ellipticals and lenticulars in SLUGGS ($\sigFJ$,
\citealt{Brodie+14}). These are the two characteristic velocities that we use to define
the Tully-Fisher and Faber-Jackson relations, which we compactly write as
\begin{equation} \label{eq:TF-FJ}
    \Mstar \propto \Vgal^a,
\end{equation}
where $\Vgal$ is $\Vflat$ for late types and $\sigFJ$ for early types.
Along either the rising branch of the SHMR for late types or along the declining one
for massive early types, the SHMR can also be approximated by a power law
\begin{equation} \label{eq:fstar-Mh}
    \fstar \propto \Mh^b.
\end{equation}
Given the definition of $\fstar$ (Eq.~\ref{eq:fstar}) and the fact that, in $\Lambda$CDM
cosmogonies, halo masses and virial velocities are related by $\Mh\propto\Vh^3$, we can
rearrange the two equations above into a relation between $\Vgal$ and $\Vh$, which
then becomes the power law
\begin{equation} \label{eq:Vgal-Vh}
    \Vgal \propto \Vh^c,
\end{equation}
with
\begin{equation} \label{eq:Vgal-Vh_exp}
    c = 3(b+1)/a.
\end{equation}
Eq.~\eqref{eq:Vgal-Vh} is key here since it relates the familiar scaling laws
(Eq.~\ref{eq:TF-FJ}) with the two branches of the SHMR (Eq.~\ref{eq:fstar-Mh}).
With our set of measurements -- $\Mstar$ from 3.6$\mu$m photometry, $\Vgal$ from
observed HI ($\Vflat$) or stellar kinematics ($\sigFJ$), $\Mh$ from dynamical models
of rotation curves and GC kinematics -- we perform power-law fits to
Eq.~\eqref{eq:TF-FJ}-\eqref{eq:Vgal-Vh}, finding
\begin{equation}
\begin{split}
    a &= 4.3 \pm 0.3,    & b = 0.5 \pm 0.2, \\
    c &= 1.02 \pm 0.21,  &\mbox{(late types)}
\end{split}
\end{equation}
\begin{equation}
\begin{split}
    a &= 3.5 \pm 0.7    & b = -0.6 \pm 0.1, \\
    c &= 0.36 \pm 0.17  &\mbox{(early types)}.
\end{split}
\end{equation}
In Fig.~\ref{fig:TF-FJ-fstar}, we show (on the top row) the data and the power-law fits
of the three relations ($\fstar-\Mh$, $\Mstar-\Vgal$, and $\Vgal-\Vh$) for the late types
(blue) and early types (red).
While the mass-velocity scaling laws have a similar slope ($a$), the SHMR and the $\Vgal-\Vh$
relations have significantly different slopes ($b$ and $c$) on the two branches. At a fixed
stellar mass, the large difference in halo mass between discs and spheroids is hidden in
the similar Tully-Fisher and Faber-Jackson relations by the different $\Vgal-\Vh$ relations.

The slope $c\approx 1$ for late types on the rising branch of the SHMR means that the
ratio $\Vflat/\Vh$ is nearly the same (and about equal to unity) for discs of all
masses \citep{Posti+19b}. This is yet another manifestation of the so-called disc-halo
conspiracy, i.e. rotation curves are observed to be flat from the inner, baryon-dominated
parts of galactic discs to the outer, dark matter-dominated parts
\citep[e.g.][]{vanAlbada+85,Kent87}. On the other hand, for early types along the falling
branch of the SHMR, we find a very different result: $c\approx 0.4$, which implies that
the ratio $\sigFJ/\Vh$ decreases with both stellar and halo mass.

While $\Vflat$ for late types is measured at large radii and traces the potential of
the dark halo, $\sigFJ$ for early types is measured in the inner regions where
the potential is dominated by stars. One might then wonder whether this is responsible
for the difference in the $\Vflat/\Vh$ and $\sigFJ/\Vh$ ratios.
We check for this in the bottom panels of Fig.~\ref{fig:TF-FJ-fstar}, where we replace
the observed $\Vgal$ with the circular velocity evaluated at a fixed radius of
$2R_{\rm e}$ for both late-type and early-type galaxies in our sample.
For late types, we obtain the circular velocities directly from the observed rotation
curves, while for early types, they are an output of our $f(\bJ)$ dynamical models.
Fig.~\ref{fig:TF-FJ-fstar} shows that the difference persists in the relation
$\Vcirc(2R_{\rm e})\propto\Vh^{\cprime}$. Now we find $\cprime\approx 1$ for late types
and $\cprime\approx 0.57$ for early types.

\subsection{Physical interpretation}

The observed scaling relations between the rotation velocities, sizes, and stellar
masses of spiral galaxies indicate that they represent a self-similar population of
objects, homologous to their dark matter halos \citep[e.g.][]{Posti+19b}. In particular,
galactic discs have almost as much specific angular momentum as their dark halos, as
expected from simple conservation arguments \citep[e.g.][]{FallEfstathiou80,Dalcanton+97,
MMW98}. Stellar feedback modifies this behaviour, making gas retention mass dependent,
and thus creating the rising branch of the SHMR. Gravitational clustering and accretion
moves galaxies along the scaling laws and up the rising branch of the SHMR. This
introduces no features in either the scaling laws or the $\fstar-\Mstar$ relation.

However, galaxies in crowded environments, such as groups and clusters,
often collide and merge with each other. Depending on the mass ratio of the galaxies
and on whether they are gas rich or gas poor, mergers have a couple of important
implications for the evolution of massive galaxies. First, the stellar body is
dynamically heated, causing the spheroidal component to grow at the expense of the
disc component, thus leading to morphological transformation
\citep[e.g.][]{Quinn+93,Hopkins+10b,Martin+18}.
Second, some of the gas in the merging galaxies may be funnelled into the central
black hole, thus triggering AGN feedback \citep[e.g.][]{Hopkins+06}. Outflows
and radiation from the AGN may then impede further inflows and star formation,
hence reducing $\fstar$. As a consequence of merging and AGN feedback, massive discs
are driven off the rising branch of the SHMR, becoming passive spheroids on the falling
branch.

In the previous subsection, we showed that late types follow a $\Vgal-\Vh$ relation with
a slope of $c\approx 1$, implying that the ratio of binding energy per unit mass of
the luminous galaxy to that of its dark halo ($\propto \Vgal^2/\Vh^2$) is approximately
independent of mass. In contrast, early types have $c\approx 0.4$, indicating that the
ratio of galaxy-to-halo binding energy per unit mass decreases as mass increases. Both
merging and AGN feedback may reduce the binding energy per unit mass, leading to $c<1$,
as observed.

To understand how mergers can lower the ratio $\Vgal/\Vh$ we consider a typical collision
between two galaxies on a weakly bound orbit \citep[seee.g.][]{KhochfarBurkert06}.
While the halo mass $\Mh$ and the virial velocity $\Vh$ increase during
merging, idealised simulations have shown that the internal velocity dispersion of
the stars typically remains constant or decreases, depending on the mass ratio, gas
fraction and orbital parameters
\citep[e.g.][]{Nipoti+03,Cox+06,Naab+09,Hilz+12,Posti+14}.
After some Gyrs of evolution in a dense environment, a massive early type that
experiences several mergers will therefore lower its $\Vgal/\Vh$ ratio as its mass
increases.
This effect is also observed in cosmological simulations, where
frequent minor mergers deposit stars primarily in the outskirts of massive galaxies,
thus lowering their binding energy per unit mass \citep[e.g.][]{Oser+12,GaborDave12}.

At the same time, outflows launched by the AGN will interact with the circumgalactic
medium, pushing some of it outward, depending on the opening angle of the outflow.
If this gas is ever able to condense and form stars, this would also tend to lower
the binding energy per unit mass of the host galaxy.

Mergers and AGN feedback may thus combine to transform star-forming discs on the rising
branch of the SHMR into quenched spheroids on the falling branch. These two processes
are contemporaneous but episodic. Both mergers and AGN feedback tend to disrupt inflow
onto galactic discs, thus suppressing disc growth while promoting spheroid growth. Between
these episodes, relatively smooth inflow can resume, thus promoting the regrowth of discs.
This reasoning suggests that massive galaxies may evolve along complicated, essentially
stochastic, paths in the region of the $\fstar-\Mstar$ plane bounded by the rising
pure-disc branch and the declining pure-spheroid branch.

\section{Summary and Conclusions} \label{sec:concl}

In this paper, we have derived the SHMR for a sample of 25 massive early-type
galaxies from estimates of their individual halo masses. We accomplished this by comparing
a dynamical model with a flexible analytical distribution function with position and velocity
data for the globular cluster systems around these galaxies. Combining our new results for
early types with those from \citetalias{PFM19} for late types based on extended HI rotation
curves, we derived, for the first time, the $\fstar-\Mstar$ relation for galaxies of different
morphologies with identical assumptions about their halo properties. Our main findings can be
summarised as follows.
\begin{itemize}
    \item[(i)] At the high-mass end of the SHMR ($\Mstar> 5\times 10^{10}\Msun$), late types
               are found to have significantly higher $\fstar$ than early types of the same
               stellar mass (by about a factor $\sim 7$ at $\Mstar\sim 10^{11}\Msun$). While
               $\fstar$ increases with $\Mstar$ for late types \citepalias{PFM19}, it decreases
               for early types, in broad agreement with expectations from abundance matching
               \citep[e.g.][] {Moster+13}. Our determinations show unequivocally that the SHMR
               has a secondary correlation with galaxy type at the high-mass end.
    \item[(ii)] For massive galaxies ($\Mstar> 5\times 10^{10}\Msun$), we studied how $\fstar$
                deviates from the expectations of abundance matching ($f_{\star,\rm AM}$) as
                a function of disc fraction and Hubble type. We find a fairly continuous
                transition between close agreement, $\log(\fstar/f_{\star,\,\rm AM})\sim 0$,
                for pure spheroids, and an order of magnitude discrepancy,
                $\log(\fstar/f_{\star,\,\rm AM})\sim 1$, for pure discs. This transition occurs
                at about $D/T \sim 0.6$, or $T\sim 2$, suggestive of scenarios involving
                merging and AGN feedback.
    \item[(iii)] We have tested the sensitivity of our $\Mh$ estimates with respect to our
                adopted priors on the $c-\Mh$ correlation.
                We find that the secondary correlation of the SHMR with galaxy type is robust
                relative to any reasonable adjustments to this prior. We have also compared
                our results both with other statistical derivations of the SHMR (e.g. using
                weak-lensing or satellite kinematics) and with other individual estimates of
                halo masses based on different data and/or techniques. We find these estimates
                to be compatible within the uncertainties, allowing us to conclude that the
                issue of whether the SHMR has a secondary correlation with galaxy type is now
                settled.
    \item[(iv)] We investigated the apparent paradox between the two separate branches of the
                 SHMR -- a rising one for discs and a falling one for massive spheroids -- and
                 the similar power-law relations between stellar masses and velocities
                 for late types and early types, the Tully-Fisher and Faber-Jackson relations.
                 We demonstrated that this happens because the relations between galaxy velocity
                 and halo velocity are different for galaxies of different types. Discs have a
                 constant ratio $\Vgal/\Vh\approx 1$ at all masses -- indicating that they are
                 close to homologous with their dark halos -- while spheroids have a declining
                 ratio $\Vgal/\Vh$ with mass. We suggest that this is a signature of the
                 combined effects of merging and AGN feedback.
\end{itemize}

As suggested above, both merging and AGN feedback are likely responsible for splitting the SHMR
and the $\Vgal-\Vh$ relation into different branches for discs and spheroids, but their exact
roles remain to be determined. The growth of discs and spheroids in massive galaxies may be
intermittent, with disc growth during periods of relatively smooth inflow, interrupted
by spheroid growth during episodes of merging and AGN feedback. Hydrodynamical simulations
may shed light on the underlying physical processes, so long as they are relatively insensitive
to numerical resolution and subgrid recipes for stellar and AGN feedback.
A careful census of black holes in a large sample of host galaxies of different morphologies
and masses likely would also be instructive.

\begin{acknowledgements}
We thank Michal B\'{i}lek, Benoit Famaey, and Filippo Fraternali for encouragement in the early
stages of this project and Romeel Dav\'{e}, Ken Freeman, and Andrey Kravtsov for helpful comments
in the later stages.
LP acknowledges support from the Centre National d’Etudes Spatiales (CNES).
This research has made use of "Aladin sky atlas" developed at CDS, Strasbourg
Observatory, France \citep{Bonnarel+00}
\end{acknowledgements}

\bibliographystyle{aa} 
\bibliography{refs} 

\appendix

\section{Action-based dynamical models of early-type galaxies} \label{app:model}

Here, we describe the dynamical models that we use to represent the distribution
function of globular cluster systems around 25 ellipticals and lenticulars and their dark
matter halos. We first summarise the basic principles of models based on action-angle variables,
and then we describe our application to the study of early types.
For a more complete introduction to action-angle variables, we refer the reader to the monographs
by \cite{Born27} and \cite{Arnold78}. We use the code \texttt{AGAMA} \citep{AGAMA} to evaluate
actions, potentials, and distribution functions, and to generate the dynamical models in this work.

\subsection{Preliminaries}

We begin with the distribution function (DF) for a globular cluster system $f$, defined such
that $f(\bx,\bv)\de\bx\de\bv$ is the probability of finding a cluster in the infinitesimal
volume element $\de\bx\de\bv$ at the position-velocity point $(\bx,\bv)$ in phase-space.
According to the strong form of the \cite{Jeans1915} theorem, in a steady state, $f$ is a function
of the integrals of motion \citep[see also][]{Lynden-Bell62}. Without loss of generality, we may
choose these to be the three action integrals
\begin{equation}
    J_i=\frac{1}{2\pi}\oint p_i\de q_i \quad \mbox{for }i=1,2,3,
\end{equation}
where $p_i$ and $q_i$ are canonically conjugate momenta and coordinates, and write the DF as
$f(\bJ)$.

Actions $\bJ$ and their canonically conjugate angles $\btheta$, are the ``natural'' coordinates
of galactic dynamics since i) the description of orbits becomes mathematically simplest,
ii) they describe systems both in and out of equilibrium and iii) actions are adiabatic
invariants, i.e. they are constant under slow changes of the potential \citep[see e.g.][]{BT08}.
$f(\bJ)$ models have been somewhat underused in galactic dynamics, mainly because actions
generally cannot be expressed with algebraic functions of positions and velocities and need to be
computed numerically. In recent years, several crucial advances have made it feasible to calculate
$\bJ$ efficiently in arbitrary potentials \citep[see e.g.][and references therein]{SandersBinney16}.
This, in turn, has led to the introduction of several analytic $f(\bJ)$ DFs tailored to model
different galaxy components \citep[e.g.][]{Binney10,Posti+15,SandersEvans15,Pascale+18,AGAMA}.

In this work, we deal mostly with spherical potentials, which greatly simplifies the
numerical calculations. In the case of a spherical system the motion of particles is confined
to a plane and all orbits can be characterised by two actions. One of these is the total
angular momentum $L=|{\bf L}|$, and the other is the radial action
\begin{equation}
    J_r = \frac{1}{\pi}\int_{r_{\rm apo}}^{r_{\rm peri}} p_r \de r =
          \frac{1}{\pi}\int_{r_{\rm apo}}^{r_{\rm peri}} \left(2E-2\Phi-L^2/r\right)^{1/2}\de r,
\end{equation}
where $p_r$ and $r$ are the radial momentum and position, $E=p_r^2/2+L^2/2r^2+\Phi$ is the energy,
$\Phi$ is the gravitational potential and $r_{\rm apo}$ and $r_{\rm peri}$ are the apocentre and
pericentre of the orbit. Thus, in the spherical case, we have $\bJ = (\Jr, L)$ and $f(\bJ)=f(\Jr,L)$.

\subsection{Distribution function} \label{app:sec:df}

To describe the phase-space distribution of globular clusters we use the DF introduced by
\citet[][see also \citealt{WilliamsEvans15}, \citealt{AGAMA}]{Posti+15}. This is
\begin{equation} \label{eq:DF}
    f({\bf J}) = \frac{M_0}{(2\pi J_0)^3} \left[1+\left(\frac{J_0}{h({\bf J})}\right)^
                                                {\rm A}\right]^{\Gamma/{\rm A}}
                 \left[1+\left(\frac{g({\bf J})}{J_0}\right)^{\rm A}\right]^{({\rm B}-\Gamma)/{\rm A}}
\end{equation}
where
\begin{equation} \label{eq:handg}
\begin{split}
    h({\bf J}) &= \nu_h\Jr + \frac{3-\nu_h}{2}L, \\
    g({\bf J}) &= \nu_g\Jr + \frac{3-\nu_g}{2}L.
\end{split}
\end{equation}
Here $M_0$ is a parameter proportional to the mass of the system described by the DF;
since we are treating the globular clusters as tracers of the potential, $M_0$ is unimportant
in this context. The DF of Eq.~\eqref{eq:DF} has 6 free parameters, all with specific
physical meanings.
$\Gamma$ and B control the asymptotic slopes of the density profile in the inner ($r\to 0$) and
outer parts ($r\to \infty$) respectively, while the parameter A controls the sharpness of the
transition between these regimes. In the case ${\rm A}=1$, the two slopes $\Gamma$ and
B have a direct correspondence to the asymptotic slopes of the density distribution\footnote{
\cite{Posti+15} showed that a self-consistent model with DF as in Eq.~\eqref{eq:DF} and
${\rm A}=1$ generates a density law that is roughly equal to that of a $\alpha\beta\gamma$-model
\citep{Zhao96} with $\alpha=1$, $\Gamma = (6-\gamma)/(4-\gamma)$ and ${\rm B} = 2\beta-3$.
}. $J_0$ is a characteristic action that defines the radial scale at which the transition
between the two regimes occurs. The last two parameters, $\nu_h$ and $\nu_g$, control the
velocity anisotropy of the model in the inner and outer parts, respectively.

An important advantage of the double power-law $f(\bJ)$ in Eq.~\eqref{eq:DF}, over models
that depend on $(E,L)$, is that in the former case the density distribution effectively
decouples from the velocity distribution.
This allows us to fix at the outset the parameters of the DF that regulate the density profile
of a GC system (A, B, $\Gamma$) and then to fit only for those that determine the velocity
anisotropy of the system ($J_0$, $\nu_h$, $\nu_g$). Such decoupling is possible because, for
double power-law models, the homogeneous functions $h(\bJ)$ and $g(\bJ)$ are designed to
approximate surfaces of constant energy in action space \citep{Williams+14,Posti+15}.
Thus, $h$ and $g$ largely determine the differential energy distribution $\de N/\de E$, and
hence the density profile of the model \citep[see S4.3 in][]{BT08}. Starting from a
quasi-ergodic model, where $\Jr$ and $L$ appear on an equal footing in $h$ and $g$, one can
easily make the model anisotropic by varying $\nu_h$ and $\nu_g$ without altering the radial
density profile \citep{Binney14,Posti+15}.

As a first step in our modelling procedure, we fix the two slopes $\Gamma$ and B and the
sharpness A by matching to the observed number density profile of GCs.
In Fig.~\ref{fig:showcase_ngc4494}b, we show this fit for the GC system of the galaxy NGC 4494,
with (A, B, $\Gamma)=(2.1, 5.3, 0.9)$. We find that, in all cases, the double power-law density
profiles generated by Eq.~\eqref{eq:DF} provide a very good description of the observed GC
number density profiles.
The remaining three parameters, $J_0$, $\nu_h$ and $\nu_g$, are instead allowed to vary,
but, for internal consistency of the DF, we need to require $0<\nu_h,\nu_g<3$
\citep[see][]{Posti+15,AGAMA}.

\subsection{Gravitational potential}

We model the mass distribution of each galaxy with two spherical components; the stellar
body of the galaxy and its dark matter halo -- the globular cluster system is then regarded
as a tracer with negligible mass.
The stellar distribution is described by a numerically deprojected \cite{Sersic68} profile,
whose parameters are taken from the photometry of 3.6 $\mu$m Spitzer images by
\cite{Forbes+17a}. We fix all the parameters of the stellar component, except its
mass-to-light ratio at 3.6 $\mu$m, which we allow to vary with a log-normal prior with a
central value estimated by \cite{Forbes+17a} from stellar population models, and a dispersion
of 0.2 dex.

The dark matter halo in our model is described by a standard NFW profile. This has two free
parameters: the virial mass ($\Mh$) and concentration ($c$), which we allow to vary.
While we adopt a flat (uninformative) prior on the halo mass, we use a prior for the
concentration that follows the mean $c-\Mh$ correlation from $\Lambda$CDM simulations, with
a scatter of 0.11 dex \citep{DuttonMaccio14}.
Thus, overall our models have six free parameters: three for the potential and three for the
DF.

Several of our galaxies appear flattened on the sky, so it is important to evaluate whether
the assumption of spherical symmetry for the stellar component of the potential significantly
affects our results. To check this, we have re-run all of our models with an axisymmetric
galaxy mass distribution that has the same 3D flattening as the 2D image \citep{Forbes+17a},
while the dark matter halo is still spherical. With respect a spherical galaxy with the same
mass, the deviations in the resulting halo masses are always well within the
uncertainties\footnote{
We also recall that the potential is always more spherical than the mass distribution that
it generates \citep{BT08}.
}.
We are therefore confident that the assumption of spherical symmetry in the galactic mass
distribution does not significantly bias our results.

\subsection{Parameters estimation}

We estimate the posterior distributions of the model parameters ($\pmb{\varpi}$) with
standard Bayesian inference: $P(\pmb{\varpi}|{\bf d}) \propto P({\bf d}|\pmb{\varpi})\,
P(\pmb{\varpi})$, where ${\bf d}$ are the data, $P({\bf d}|\pmb{\varpi})$ is the likelihood,
and $P(\pmb{\varpi})$ is the prior.
We adopt a prior that is flat (uninformative) for four parameters ($\log\,\Mh$, $\log\,J_0$,
$\nu_h$ and $\nu_g$), gaussian for $\log\,M/L_{3.6}$, with a mean estimated for each galaxy
by \cite{Forbes+17a} with stellar population models and a dispersion of 0.2 dex, and gaussian
for $\log\,c$, with a mean given by the $\Lambda$CDM relation and a dispersion of 0.11 dex.

The DF in Eq.~\eqref{eq:DF} itself is a probability distribution that can serve as the
likelihood in our framework. Specifically, for a set of $N$ particles with position-velocity
coordinates $(x_i,v_i)$ orbiting in a given potential $\Phi$, the likelihood, given the model
$f({\bf J})$, is simply $\prod_{i=0}^N f[{\bf J}(x_i,v_i)]$. In reality, when dealing with data,
one does not know the positions and velocities with infinite precision; thus, a convolution of
the DF with the observed error distribution is needed \citep[see][]{BinneyWong17,PostiHelmi19}.

In our case, we also lack information about the two transverse velocities and the precise
positions of the GCs along the line-of-sight (LOS). To take this into account, we marginalise
the likelihood over all of the realistically possible transverse velocities and LOS positions
of the clusters. For the two unknown transverse velocities $(v_x,v_y)$, we adopt uniform
distributions in the range $[-\Vesc,\Vesc]$, where $\Vesc$ is the escape velocity of the
potential. For the unknown LOS position $z$, we adopt the deprojected density distribution
of the GC system $\rho(s)$, where $s$ is the spherical radius,
$s^2=x_{\rm GC}^2+y_{\rm GC}^2+z^2$, evaluated at a fixed position on the sky
$(x_{\rm GC}, y_{\rm GC})$. Thus, we have the following error distribution

\begin{equation} \label{eq:Edistrib}
\begin{split}
    \mathcal{E}({\bf u}|{\bf d}) = &\delta(x-x_{\rm GC})\,\delta(y-y_{\rm GC})\,
                                     G(v_z|\Vrad,\epsilon_{\Vrad})\\
                                   &\rho(z)\,U(v_x|-\Vesc,\Vesc)\,U(v_y|-\Vesc,\Vesc),
\end{split}
\end{equation}
where ${\bf u}=(x,y,z,v_x,v_y,v_z)$ is a point in phase space in a Cartesian frame centred
on the galaxy, and ${\bf d}=(x_{\rm GC}, y_{\rm GC}, \Vrad, \epsilon_{\Vrad})$ are the
observations. Here $\rho$ is the deprojected density distribution derived from the observed
GC number counts profile (Fig.~\ref{fig:showcase_ngc4494}b), $G(v_z|\Vrad,\epsilon_{\Vrad})$
is a gaussian distribution with mean $\Vrad$ and dispersion $\epsilon_{\Vrad}$,
$U(v|-\Vesc,\Vesc)$ is a uniform distribution in the range $[-\Vesc,\Vesc]$, and
$\delta(x-x_{\rm GC})$ is a Dirac delta distribution centred on $x_{\rm GC}$. We use a $\delta$
distribution because the uncertainty in the sky positions of the clusters is negligible.
Finally, the likelihood of our model is given by the convolution of the DF with the
$\mathcal{E}$ distribution of each cluster, i.e.

\begin{equation} \label{eq:like}
    P({\bf d}|\pmb{\varpi}) = \prod_{i=0}^N \int {\rm d}{\bf u}\,\mathcal{E}({\bf u}|
                               {\bf d}_i)\,f[{\bf J}({\bf u})].
\end{equation}

In practice, we evaluate Eq.~\eqref{eq:like} with a Monte Carlo method, sampling the integral
and the $\mathcal{E}$ distribution of each cluster with 1000 realisations. Fortunately, the
likelihood in Eq.~\eqref{eq:like} turns out to be quite insensitive to the specific form of
both the density distribution $\rho$ and the distributions of the missing velocities; in
fact, we verified that using a gaussian instead of a uniform distribution in $v_x$ and $v_y$
does not alter significantly our results on the halo masses.
As an example, in Fig.~\ref{fig:showcase_ngc4494}d, we show the distribution of clusters around
NGC 4494 on the observable projection of the phase-space, the $\Vlos-r$ plane, compared to the
prediction of the maximum-likelihood $f({\bJ})$ model for this galaxy.
The $\Vlos-r$ plane is effectively the sub-space where we are fitting our models to the data.

With the prior and likelihood defined as above, we evaluate the posterior distribution of the
six free parameters of the model with a Markov Chain Monte Carlo (MCMC) method; in particular,
we use the affine-invariant sampler implemented in the code \texttt{emcee} by \cite{emcee}.
For all 25 SLUGGS early-type galaxies, we find that the chains converge quite rapidly
around a well-defined peak in the posterior after a short burn-in phase. As an example, in
Fig.~\ref{fig:showcase_ngc4494}c, we show the marginalised posterior distributions for the
halo mass and concentration for the galaxy NGC 4494. Clearly, both $\Mh$ and $c$ are well
constrained by our analysis, despite having an unavoidable degeneracy. For each
parameter, we take the median of the marginalised posterior as the best-fit value and the
interval between the 16th and 84th percentiles as a measure of its uncertainty.

\subsection{Derived quantities}

From our model, with parameters optimised for each GC system in the SLUGGS sample, we can now
derive several other properties of interest. In Fig.~\ref{fig:showcase_ngc4494}e, we show, as
examples, the circular velocity curve of the mass distribution, $\Vcirc$, and the (spherically
averaged) velocity anisotropy profile of the GC system,
$\beta = 1 - (\sigma_\theta^2+\sigma_\phi^2)/2\sigma_r^2$.
While $\Vcirc$ depends only on the three free parameters of the gravitational
potential $(\Mh,c,M/L_{3.6})$, $\beta$ depends mostly on the three free parameters of the DF
($\nu_h, \nu_g, J_0$). This means, incidentally, that the uncertainty on $\Vcirc$, which
we estimate with random realisations of the model from the posterior ($1\sigma$ grey band in
Fig.~\ref{fig:showcase_ngc4494}e), is fully determined by the width of the posterior on the
parameters of the potential (Fig.~\ref{fig:showcase_ngc4494}c).

We can also compute the profile of the LOS velocity dispersion of the GC systems ($\sigmalos$),
which depends on both the potential and the DF. We show this model profile for NGC 4494 in
Fig.~\ref{fig:showcase_ngc4494}f, where we compare it with the observed profile
\citep[see][]{Foster+16}.
Such a comparison is meaningful since we do not input directly the $\sigmalos$ profile to
our fitting routine, although we do, of course, input the same individual velocities that
determine $\sigmalos$. The agreement that we observe for NGC 4494
(Fig.~\ref{fig:showcase_ngc4494}f), and also for the other galaxies in our sample (not shown),
thus serves as a useful consistency check on our procedure.

For 18 of the 25 early types in our sample, \cite{Pulsoni+18} measured the velocity dispersion
profile of the population of planetary nebulae orbiting around the host galaxy and found that
in most cases it agrees quite well with the $\sigmalos$ profile of the GC system from SLUGGS.
Fig.~\ref{fig:showcase_ngc4494}f shows this agreement for NGC 4494. In a few cases, however,
the $\sigmalos$ for the planetary nebulae is $\sim 20-40\%$ lower than for the globular
clusters. This difference in the $\sigmalos$ likely reflects the different density profiles
of the two types of tracers orbiting in the same gravitational potential. Even among GCs,
there are differences in $\sigmalos$ when the system is subdivided by colour.
Red GCs have lower $\sigmalos$ than blue GCs, and are in better agreement with both the
velocity dispersion of planetary nebulae and the stellar bodies of galaxies
\citep[e.g.][]{Pota+15}.

\end{document}